\pdfoutput=1

\documentclass[12pt,preprint]{aastex}

\usepackage{natbib}

\renewcommand{\vec}[1]{ {\mathbf #1} }

\newcommand{\jcphy}{{J.~Comput.~Phys.}}

\newcommand{\hlenfig}{0.45\textwidth}

\newcommand{\Eq}{{Equation}}

\newcommand{\Fig}{{Figure}}
\newcommand{\Figs}{{Figures}}

\newcommand{\dive}{\nabla\cdot}
\newcommand{\divB}{\nabla\cdot\mathbf{B}}
\newcommand{\crlB}{\nabla\times\mathbf{B}}

\shorttitle{New Implementation of NLFFF Extrapolation}
\shortauthors{Jiang et al.}

\begin{document}


\title{A New Implementation of the Magnetohydrodynamics-Relaxation
  Method for Nonlinear Force-Free Field Extrapolation in the Solar
  Corona}


\author{Chaowei Jiang and Xueshang Feng} 
\affil{SIGMA Weather Group, State Key Laboratory for Space Weather,
  Center for Space Science and Applied Research, Chinese Academy of
  Sciences, Beijing 100190} 
\email{cwjiang@spaceweather.ac.cn; fengx@spaceweather.ac.cn}




\begin{abstract}
  Magnetic field in the solar corona is usually extrapolated from
  photospheric vector magnetogram using a nonlinear force-free field
  (NLFFF) model. NLFFF extrapolation needs a considerable effort to be
  devoted for its numerical realization. In this paper we present a
  new implementation of the magnetohydrodynamics (MHD)-relaxation
  method for NLFFF extrapolation. The magneto-frictional approach
  which is introduced for speeding the relaxation of the MHD system is
  novelly realized by the spacetime conservation-element and
  solution-element (CESE) scheme. A magnetic field splitting method is
  used to further improve the computational accuracy. The bottom
  boundary condition is prescribed by changing the transverse field
  incrementally to match the magnetogram, and all other artificial
  boundaries of the computational box are simply fixed. We examine the
  code by two types of NLFFF benchmark tests, the \citet{Low1990}
  semi-analytic force-free solutions and a more realistic solar-like
  case constructed by \citet{Ballegooijen2007}. The results show that
  our implementation are successful and versatile for extrapolations
  of either the relatively simple cases or the rather complex cases
  which need significant rebuilding of the magnetic topology, e.g., a
  flux rope. We also compute a suite of metrics to quantitatively
  analyze the results and demonstrate that the performance of our code
  in extrapolation accuracy basically reaches the same level of the
  present best-performing code, e.g., that developed by
  \citet{Wiegelmann2004}.
\end{abstract}


\keywords{Magnetic fields; Magnetohydrodynamics (MHD); Methods:
  numerical; Sun: corona}

\section{Introduction}
\label{sec:1}

The magnetic field configuration is essential for us to understand the
solar explosive phenomena, such as flares and coronal mass
ejections. Besides, the magnetic field also plays a crucial role in
determining the slowly-evolving structures of the solar corona, such
as the coronal streamers and the coronal holes. However, direct
measurements of these magnetic fields are very difficult to implement,
and the present observations for the magnetic fields based on the
spectropolarimetric method (the Zeeman effect and the Hanle effect)
are basically restricted on the visible surface layer, i.e., the
photosphere. Even a routine recording of the full surface fields on
the photosphere are only available for the line-of-sight (LoS)
component (e.g., the daily disk magnetogram provided by
SOHO/MDI). Most of the vector magnetograms at present are
recorded locally for active regions 
and some of them may be unreliable because of large random errors and
the $180^{\circ}$ ambiguity. In view of these limitations, researchers
hence resort to using physical models to extrapolate (or reconstruct)
the coronal fields from the observable photospheric magnetogram
\citep{Sakurai1989,Aly1989,Amari1997,McClymont1997,Aschwanden2005,Wiegelmann2008}.

On large scale with relatively low resolutions, the corona fields are
usually extrapolated from the LoS magnetogram with models including
the potential field source surface model
\citep{Altschuler1969,Hoeksema1984} and the MHD models
\citep{Mikic1999,Linker1999,Feng2007,Feng2010}. By these models and
the global map of photospheric field, i.e., the synoptic map as
usually called, the extrapolated global fields can be used to study
the general structures of the corona and the heliosphere (e.g., the
locations, shapes and sizes of coronal holes, coronal streamers and
heliospheric current sheet and their evolutions). On the local scale
with high resolutions, when one's interest is focused on the active
regions, a most common and powerful way of reconstructing the magnetic
fields is the nonlinear force-free field (NLFFF) extrapolation from
the vector magnetogram. The force-free assumption is a good
approximation for fields in the low corona but above the
photosphere. It is because in most parts of the low corona,
particularly the strongly-magnetized active regions, the plasma
$\beta$ (a ratio of gas pressure $p$ to magnetic pressure
$B^{2}/(2\mu_{0})$, i.e., $\beta=2\mu_{0}p/B^{2}$) is extremely low
($\beta \ll 1$) and the plasma velocity $v$ is also low compared to
the Alfv\'en speed $v_{A}$ ($v \ll v_{A}$), which means that the
pressure gradient, gravity, and inertial force can be neglected from
the momentum equation and thus the only-survived Lorentz force must be
self-balanced, i.e., $\vec j\times \vec B = \vec 0$ ($\vec j$ is the
electric current density and $\vec B$ is the magnetic field). This
means that $\crlB = \alpha\vec B$, where the scalar $\alpha$ is called
the force-free parameter. Generally, $\alpha$ varies spatially for
NLFFF and some popular simplifications include $\alpha=0$ for
potential field and $\alpha={\rm constant}$ for linear force-free
field.

The reasons why nonlinear force-free model is superior over other much
simpler force-free models for the active regions, i.e., the potential
field and the linear force-free field models are mainly as follows
\citep{Wiegelmann2008}: (1) observation shows that there are
significant non-potential fields in the active regions, which excludes
the potential model; (2) usually the force-free parameter $\alpha$ is
a very space-dependent function as derived from the measured vector
magnetogram and also demonstrated by great contradiction of the
observed loops and linear force-free extrapolations; (3) potential and
linear force-free fields are too simple to estimate the free magnetic
energy and magnetic topology accurately. On the other hand, one may
wonder why the more realistic model, the MHD model (e.g.,
\citep{Wu2006,Wu2009}), is less commonly used than the NLFFF
model. The reason is twofold. Firstly, there is a lack of observed
information of gas parameters such as the surface plasma density and
velocity, which are critical boundary conditions for the full MHD
simulations \citep{Abbett2003,Abbett2004,Welsch2004,Wang2011};
secondly the numerical realization of full MHD simulation is greatly
limited by the present computational capability. For instance, the
Courant-Friedrichs-Lewy (CFL) condition puts rather severe
restrictions on the size of the time step for most explicit schemes
because densities in the corona are very low while magnetic field
strengths in active regions can be quite high ($\sim$kG) which
inescapablely results in a extremely high Alfv\'en speed
\citep{Abbett2003}. The computational limitation of full MHD arises
especially when applying to very high-resolution and large-field-view
magnetograms currently available.

A variety of numerical codes with different methods have been proposed
to implement the NLFFF extrapolations up to the present. The
underlying methods of these codes can be classified into six types
including (1) the Grad-Rubin method
\citep{Grad1958,Sakurai1981,Amari1999,Amari2006,Wheatland2004,Wheatland2006};
(2) the upward integration method
\citep{Nakagawa1974,Wu1990,Song2006}; (3) the MHD relaxation method
\citep{Chodura1981,Yang1986,Mikic1994a,Roumeliotis1996,Valori2005,Valori2007,Jiang2011,Inoue2011};
(4) the optimization approach
\citep{Wheatland2000,Wiegelmann2004,Inhester2006,Wiegelmann2006,Wiegelmann2007};
(5) the boundary element method \citep{Yan2000,Yan2006,He2008,He2011}
and (6) the most recently arisen force-free electrodynamics method
\citep{Contopoulos2011}. The reader is referred to
\citep{Wiegelmann2008} for a comprehensive review of most of these
methods. In addition to the difference in methods, the specific
realizations (i.e., the codes) differ significantly in many other
aspects from software to hardware, e.g., the mesh configuration, the
numerical scheme and boundary conditions, the language of the code
(i.e., IDL, C or Fortran), the hardware architecture and the degree of
parallelization. As a consequence, these codes perform very
differently from each other with the computational speed and
extrapolation accuracy. \citet{Schrijver2006} and \citet{Metcalf2008}
have carried out detailed comparisons of some representative codes
using respectively the semi-analytic Low \& Lou's force-free solutions
\citep{Low1990} and a Sun-like test case constructed by
\citet{Ballegooijen2007}. They show that although all the tested codes
can achieve the reference solutions qualitatively, the differences are
considerable by quantity. It is pointed out by their analysis that the
way of implementation of the method plays the same important role as
its underlying approach for causing such differences. In particular,
they found that the optimization method coded by
\citet{Wiegelmann2004} is the fastest converging and best-performing
algorithm.

In our previous work \citep{Jiang2011}, we have used a full-MHD
relaxation method for reconstructing the corona field basing on our
CESE-MHD code \citep{Feng2007,Jiang2010}. We included both the gravity
and gas pressure of plasma in the model for a more realistic emulation
of the low corona. The relaxation is solely dependent on a relatively
small viscosity term $\nabla\cdot(\nu\rho\nabla\mathbf{v})$ ($\rho$,
$\vec v$ are respectively the plasma density and velocity and $\nu$ is
the viscosity) as done by \citep{Mikic1994a}. It is demonstrated that
the MHD relaxation method combined with the established CESE-MHD code
can gain many advantages over other approaches, such as the simplicity
of the implementation, the high accuracy of the computation, and the
efficiency of the highly parallelized code. By the Low \& Lou's
force-free benchmark, it is also found that our implementation
reconstructed a result comparable with the best one by
\citet{Wiegelmann2004} reported in \citep{Schrijver2006}, which
encourages us for the further work of our method for more realistic
applications. As another fact, it also proved that the force-free
model is a good assumption since we directly start from the full MHD
and finally reach a state very force-free. Recently in the experiment
with realistic magnetogram, however, we found that the system is prone
to produce very large velocity (of several $v_{A}$) when performed on
magnetograms with rather large gradient, which is ordinary in
realistic photospheric data. It is because the discretization errors
of the large-gradient regions can cause large Lorentz forces whereas
the used viscosity $\nu$ is too small to effectively control the
motion driven by these forces. This velocity thus severely restricts
the timestep and further decreases the relaxation speed of the whole
system. On the other hand, increase of the viscosity may be useful for
restricting the plasma velocity, but it also significantly restricts
the timestep because the CFL condition says
\begin{equation}
  \Delta t < 0.5\frac{\Delta x^{2}}{\nu}
\end{equation}
where $\Delta x$ and $\Delta t$ are the mesh spacing in space and
time. Too small timestep is in particular unfavorable for the CESE
scheme, which can produce excessive numerical diffusion and lose the
accuracy \citep{Chang2002,Feng2010}. Although an implicit dealing of
this viscosity term can remarkablely remedy this problem, the price is
a big complication of the numerical scheme and parallelization.
Moreover, for the case of force-free extrapolation in which only the
magnetic field is solved, it obviously gives no payoff if considering
the computational restrictions of the full MHD model, e.g., the slow
evolution of the weak field and the additional computational resources
consumed by solving plasma density and pressure.

In this paper, we propose to present a new implementation of the MHD
relaxation method for NLFFF extrapolation to avoid the above
shortcomings of our previous method. We now adopt the
magneto-frictional approach as used by
\citep{Roumeliotis1996,Valori2007}, which explicitly introduce a
frictional-like force $\vec F=-\nu\vec v$ to the momentum equation. By
adjusting the frictional parameter $\nu$, one can control the
relaxation of the system more efficiently than using the
viscosity. Different from the convectional form of the
magneto-frictional equation as used in
\citep{Roumeliotis1996,Ballegooijen2000,Valori2007} which can not be
solved by many modern CFD (computational fluid dynamics) or MHD
solvers designed for the standard PDE (partial differential equation)
system like
\begin{equation}
  \frac{\partial\mathbf{U}}{\partial
    t}+\frac{\partial\mathbf{F}}{\partial
    x}+\frac{\partial\mathbf{G}}{\partial
    y}+\frac{\partial\mathbf{H}}{\partial z} = \mathbf{S},
\end{equation}
we use a form with the time-dependent term of momentum reserved. It
will be shown that by such modification, the equation system can be
still written in standard conservation form with source terms, for
which the CESE-MHD method is just designed. This paper also focuses on
a comprehensive examination of the implementation, by applying the
code to extrapolations of the semi-analytic force-free solutions
adopted by \citet{Schrijver2006} (hereafter referred to as Paper I)
and the more stringent solar-like test used by \citet{Metcalf2008}
(hereafter referred to as Paper II). All these tests will be carried
out with the same conditions as much as possible (i.e., the same mesh
resolution, the same initial potential field, the same artificial
boundary conditions) as in the above two papers for a rigid assessment
and comparison with reported results there. We will show that by this
new implementation, we successfully improved our method over the
previous work of \citep{Jiang2011}. Quantitative comparisons of the
results will demonstrate that our performance of the extrapolation
accuracy basically reaches the same level of the present
best-performing code by \citet{Wiegelmann2004} even for the rather
stringent test cases.

The remainder of the paper is organized as follows. In
Section~\ref{sec:method}, we describe the model equations and the
numerical implementation. In Section~\ref{sec:ref_models} we give a
briefly review of the benchmark models used for testing the code. The
metrics that is used to evaluate the results of extrapolations are
given in Section~\ref{sec:metrics} and the extrapolation results and
comparisons are reported and discussed in
Section~\ref{sec:results}. Finally, we offer concluding remarks and
some outlooks in Section~\ref{sec:conclusions}.

\section{The Method}
\label{sec:method}

\subsection{The Magneto-Frictional Equations}
\label{sec:m.1}
In the magneto-frictional method, an artificial frictional force is
introduced to the MHD momentum balance equation
\begin{equation}
  \label{eq:m.1}
  \rho\frac{D\mathbf{v}}{D t} = 
  \frac{\partial(\rho\mathbf{v})}{\partial t}+
  \nabla\cdot(\rho\mathbf{vv}) =
  -\nabla p+\rho\mathbf{g}+\mathbf{J}\times\mathbf{B}-\nu\mathbf{v}
\end{equation}
where current $\vec J=\crlB$ and $\nu$ is the frictional
coefficient. In situations for seeking a force-free field, the plasma
pressure and gravity can be neglected, which leads to the following
zero-beta equation
\begin{equation}
  \label{eq:m.2}
  \rho\frac{D\mathbf{v}}{D t} = 
  \mathbf{J}\times\mathbf{B}-\nu\mathbf{v}.
\end{equation}
By further discarding the inertial term, (i.e., ${D\mathbf{v}}/{D
  t}=\mathbf{0}$), it finally gives the usually adopted form of the
magneto-frictional method \citep{Ballegooijen2000,Valori2007},
\begin{equation}
  \label{eq:m.3}
  \nu\mathbf{v} = \mathbf{J}\times\mathbf{B}.
\end{equation}
This is simply a balance between the Lorentz force and the friction
term and thus the the velocity can be explicitly obtained in terms of
the magnetic field. This velocity from \Eq~(\ref{eq:m.3}) can then be
input to the magnetic induction equation
\begin{equation}
  \label{eq:m.4}
  \frac{\partial\mathbf{B}}{\partial t} = 
  \nabla\times(\mathbf{v}\times\mathbf{B})
\end{equation}
which drives the evolution of the magnetic field. Note that in such
simplification, the only equation that needs to be solved is the
induction equation and many simple finite-difference method can be
used to solve it, as long as the frictional coefficient is large
enough to suppress the potential numerical instability.

In this paper, we do not use the conventional form of the
magneto-frictional equation. For convenience of utilizing the existing
CESE solver, we partially reserve the inertial term in
\Eq~(\ref{eq:m.2}). Specifically, the time-dependent form of the
momentum equation are retained as follows:
\begin{eqnarray}
  \label{eq:m.5}
  \frac{\partial(\rho\mathbf{v})}{\partial t} =
  (\crlB)\times\mathbf{B}-\nu\rho\mathbf{v},
  \    
  \rho = |\mathbf{B}|^{2}+\rho_{0},
\end{eqnarray}
where only term $\nabla\cdot(\rho\mathbf{vv})$ are omitted from
\Eq~(\ref{eq:m.2}). Here the density $\rho$ is set as for an nearly
uniform Alfv\'en speed to roughly equalize the speed of evolution of
the whole field, and a small value $\rho_{0}$, e.g., $\rho_{0}=0.01$
is necessary to deal with very weak field associated with the magnetic
null. This form is also different from the zero-beta model since the
time-variation of momentum $\mathbf{p}=\rho\mathbf{v}$ is only induced
by the Lorentz force and the frictional term locally without affected
by the neighboring plasma. This benefits in the context of using a
rather non-uniform density as $\rho \propto B^{2}$.

For the induction equation, we use
\begin{equation}
  \label{eq:m.5.1}
  \frac{\partial\mathbf{B}}{\partial t} = 
  \nabla\times(\mathbf{v}\times\mathbf{B})
  -\mathbf{v}\divB+\nabla(\mu\divB).
\end{equation}
The terms $-\mathbf{v}\divB$ and $\nabla(\mu\dive\mathbf{B})$ added on
to the induction equation are both aimed for control the numerical
error of $\divB$. The first term $-\mathbf{v}\divB$ is derived from
the Powell's eight-wave MHD model \citep{Powell1999} and the second
term is a diffusive control of $\divB$ \citep{Marder1987,Dedner2002}
with diffusive coefficient $\mu$. The effect of these control terms
can be explicitly seen by take divergence of the induction equation:
\begin{equation}
  \label{eq:m.6}
  \frac{\partial\rho_{m}}{\partial t} = 
  -\nabla\cdot(\mathbf{v}\rho_{m})+\nabla^{2}(\mu\rho_{m}),\ 
  \rho_{m} = \divB.
\end{equation}
\Eq~(\ref{eq:m.6}) shows that the numerical magnetic monopoles
$\rho_{m}$, once arise (either because of the numerical error or from
the boundary conditions), can not accumulate locally. Instead, they
are convected effectively with the velocity of the plasma $\mathbf{v}$,
and meanwhile is diffused among the computational volume with speed of
$\mu$.

Another modification is made by utilizing the so-called magnetic field
splitting form of the MHD equation originated by
\citet{Tanaka1994}. By dividing the full magnetic field $\mathbf{B}$
into two parts ($\mathbf{B}=\mathbf{B}_{0}+\mathbf{B}_{1}$), a
embedded constant field $\mathbf{B}_{0}$ and a deviation
$\mathbf{B}_{1}$, accuracy can be gained by solving only the
deviation. The magnetic splitting form is usually used for the global
simulation of the solar wind or its interaction with a magnetized
planet such as earth \citep{Tanaka1994,Nakamizo2009,Feng2010}, since
in these cases a strong `intrinsic' potential magnetic field is
present. In the case of solving a force-free field, a potential field
that matches the normal component of the magnetogram can be regarded as
$\mathbf{B}_{0}$, which is only induced by the current system below
the bottom (i.e., the photosphere). While the deviation $\vec B_{1}$
can be seen as the field only induced by the currents in the
extrapolation volume (above the photosphere). Then the magnetic
splitting form of magneto-frictional method for solving NLFFF reads as
in a complete system
\begin{eqnarray}
  \label{eq:m.7}
  \frac{\partial\rho\mathbf{v}}{\partial t} =
  (\crlB_{1})\times\mathbf{B}-\nu\rho\mathbf{v},
  \nonumber\\
  \frac{\partial\mathbf{B}_{1}}{\partial t} = 
  \nabla\times(\mathbf{v}\times\mathbf{B})
  +\nabla(\mu\divB_{1})
  -\mathbf{v}\divB_{1},\nonumber\\
  \frac{\partial\mathbf{B}_{0}}{\partial t} = \mathbf{0},
  \crlB_{0}=\mathbf{0},
  \divB_{0}=0,\nonumber\\
  \rho=|\vec B|^{2}+\rho_{0},\
  \mathbf{B}=\mathbf{B}_{0}+\mathbf{B}_{1}.
\end{eqnarray}
A notable advantage of using the above equations is that we can
totally avoid the random numerical currents and divergences remaining
in the initial potential field that is computed by the Green's
function method or other numerical realization. It is commonly made
that in the extrapolation box the currents are concentrated in the
interior of the volume while the upper and the surrounding region are
dominated by the relatively weak potential field
\citep{Schrijver2008,Derosa2009}. Thus the splitting form can retain
the accuracy of this field. Other merits of using the splitting
equations will be seen in the implementation of a multigrid-type
optimization (Section~\ref{sec:m.2}).

\subsection{Numerical Implementation}
\label{sec:m.2}

The above equation system~(\ref{eq:m.7}) can be written in a general
conservation form with source terms as follows
\begin{equation}
  \label{eq:main}
  \frac{\partial\mathbf{U}}{\partial t}+
  \frac{\partial\mathbf{F}}{\partial x}+
  \frac{\partial\mathbf{G}}{\partial y}+
  \frac{\partial\mathbf{H}}{\partial z}
  -\frac{\partial\mathbf{F}_{\nu}}{\partial x}
  -\frac{\partial\mathbf{G}_{\nu}}{\partial y}
  -\frac{\partial\mathbf{H}_{\nu}}{\partial z}
  = \mathbf{S}
\end{equation}
with $\mathbf{U} =
\left(\rho\mathbf{v},\mathbf{B}_{1},\mathbf{B}_{0}\right)$ and other
terms are given in Appendix. We then input this model equations to the
CESE code, which is designed for any equations that can be written in
the above standard form. The CESE method deals with the 3D governing
equations in a substantially different way unlike the traditional
numerical methods (e.g., the finite-difference or finite-volume
schemes). The key principle also a conceptual leap of the CESE method
is to treat space and time unitedly as one entity. By introducing the
conservation elements (CEs) and solution elements (SEs) as the
vehicles for calculating space-time flux, the CESE method can enforce
the conservation laws both locally and globally in their natural
space-time unity form. Compared to many other numerical schemes, the
CESE method can achieve higher accuracy with the same mesh resolution
meanwhile provides simple mathematics and coding such as free of any
type of Riemann solver or eigendecomposition. Thus it can benefit for
the non-hyperbolic system like the present form of magneto-frictional
model (\ref{eq:m.7}). For more detailed descriptions of the CESE
method for MHD simulations, please see
\citep{Feng2006,Zhang2006,Feng2007,Jiang2010,Jiang2011s}.

The initial and boundary conditions are given as usually as done in
the MHD relaxation method for NLFFF extrapolation
\citep{Roumeliotis1996,Valori2007,Jiang2011}. The initial magnetic
field is supplied with the potential field $\vec B_{\rm pot}$ computed
using the LoS magnetogram. In the magnetic splitting form, it is
simply set by $\vec B_{0} = \vec B_{\rm pot}$ and $\vec B_{1}=\vec
0$. The system is started from a static state ($\vec v=\vec 0$) and
driven by inputting vector-magnetic information on the bottom
boundary. Specifically, at the bottom boundary, the magnetic field
$\vec B_{1}$ is changed linearly from the initial value $\vec 0$ to
the final value $\vec B_{\rm vec}-\vec B_{\rm pot}$ ($\vec B_{\rm
  vec}$ is the vector magnetogram) in several Alfv\`en time
$\tau_{A}$. In such process, the Lorentz forces are continuously
injected from the bottom to drive the system away from the initial
potential field. After then the bottom boundary is fixed for the
system to relax to a new equilibrium. During the whole evolution, the
lateral and top faces are fixed as $\vec B_{1} = \vec 0$ and the
velocity of all boundaries is unchanged as $\vec v=\vec 0$.
 
The time step $\Delta t$ is restricted by the CFL condition as
\begin{equation}
  \Delta t = 0.5\frac{\Delta x}{v_{A}+v_{\max}}
\end{equation}
where the Alfv\`en speed $v_{A}=1$ and $v_{\max}$ is the maximum
velocity of the whole computational domain. In this context, an
arbitrary choice of the frictional coefficient ($\nu > 0$) can be
workable because the numerical instability is prohibited by the CFL
condition. But choosing a proper $\nu$ is particularly important since
it controls the relaxation speed of the system. A simple
half-discretizing of the momentum equation~(\ref{eq:m.5}) gives
\begin{equation}
  \frac{\vec p^{n+1}-\vec p^{n}}{\Delta t} = \vec J\times \vec
  B-\nu\vec p^{n+1}
\end{equation}
where $n$ denotes the time level (note that the source terms, e.g.,
the friction, are treated implicitly in the CESE method), and thus
\begin{equation}
  \label{eq:m.2.1}
  \vec p^{n+1} = \frac{\vec p^{n}+\Delta t\vec J\times\vec
    B}{1+\nu\Delta t}.
\end{equation}
\Eq~(\ref{eq:m.2.1}) shows that the effect of the friction is simply
to reduce the momentum every time-step by a factor of $1+\nu\Delta
t$. A too small $\nu$ is prone to lead to a too large velocity ($\ge
v_{A}$), which may distort the field line excessively. On the other
hand, a too strong friction will suppress the velocity to a very low
value that makes the system too hard to to be driven. For a compromise
we set $\nu = 4c\Delta t/\Delta x^{2}$ which gives the factor
\begin{equation}
  1+\nu\Delta t = 1+\frac{c}{(1+v_{\max})^{2}}.
\end{equation}
where $c\sim 1$ is variable for optimizing the relaxation. In this
form, the friction is adaptively optimized for both the driving and
relaxing processes according to the maximum velocity: in the driving
process, the velocity is relatively large which thus reduces the
friction for fast evolution away from the initial field; in the
relaxing process, the velocity becomes smaller which will increase the
friction for fast relaxation to equilibrium. Similar setting of $\nu$
is also done by \citep{Ballegooijen2000,Valori2007}. Finally for the
diffusive coefficient of $\divB$, we set $\mu=0.4\Delta x^{2}/\Delta
t$ to maximize the diffusive effect without introducing numerical
instability.

One great challenge of the NLFFF reconstructions is the limitation of
computational resources, especially for the extrapolation of currently
available high-resolution and large-field-view magnetograms, thus a
parallel computation is generally necessary to be used. Our method is
parallelized by the AMR-CESE code \citep{Jiang2010,Jiang2011} which is
a combination of the CESE code within the PARAMESH toolkit (a
open-source Fortran package for implementation of the parallel-AMR
technique on existing code \citep{MacNeice2000}), and is performed on
a share-memory parallel cluster. Also for a large magnetogram, a
multigrid-like strategy is recommended to be used to both accelerate
the computation and improve the quality of extrapolation
\citep{Metcalf2008}. We compute the solution serially on a number of
grids with the resolution ratio of two, and input the results of the
coarser resolution to initialize the next finer resolution. It should
be noted that such method is not standard multigrid, since it does not
incorporate different grids simultaneously and iterate back and forth
between coarser and finer grids, but computes the solution of
different grids only once with the coarser solution used to initialize
the finer grid. The main advantage by doing this is to give a better
(than potential field) starting equilibrium on the full resolution
grid. Particularly standard node-centered full-weighting restriction
and prolongation operators of multigrid method are used to transfer
data between different resolutions. By these operators, any boundary
values are interpolated using data on the same boundary face and the
total flux of the magnetogram is conserved between different grids. It
is worthwhile noting that when using the coarser results to initialize
the finer grid, the magnetic splitting form can gain accuracy since
only the deviation field $\vec B_{1}$ is needed to interpolate, while
the intrinsic potential field $\vec B_{0}$ is always reset by the
value on the final resolution or the Green's function method. However,
the multigrid algorithm presents demerit because every prolongation
(by interpolation) will introduce new errors of divergence and current
to $\vec B_{1}$ which can be `felt' by the CESE method. Such problem
is also faced in many AMR simulations due to the mesh refinement
\citep{Toth2002}. We will discuss its effects in
Section~\ref{sec:results} by comparing the results with and without
the multigrid algorithm.

\section{Benchmark Models}
\label{sec:ref_models}

\subsection{Low and Lou Force Free Field}
\label{sec:ref.1}

\begin{figure}[\htbp]
  \centering
  \includegraphics[width=\textwidth]{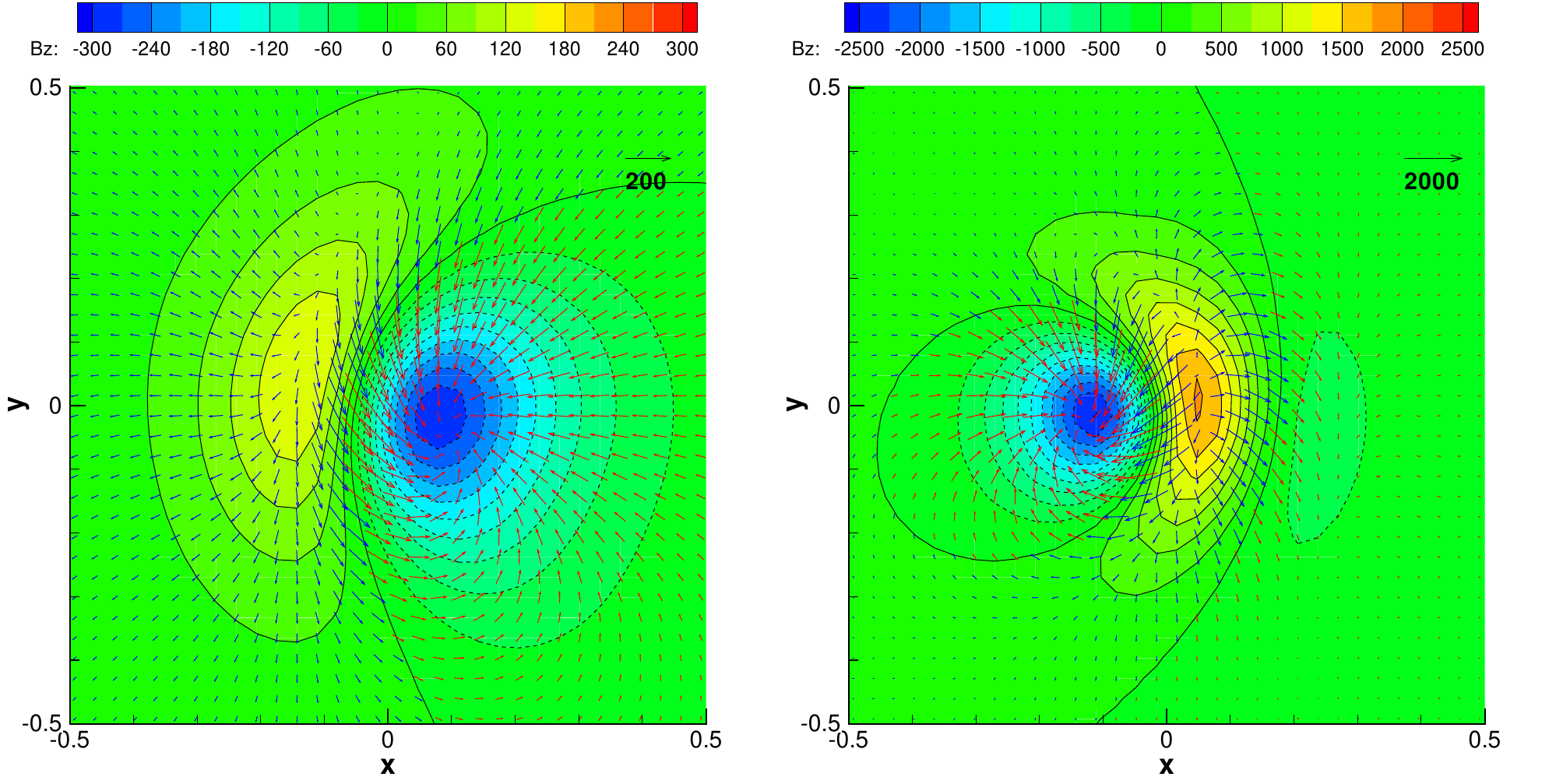}
  \caption{Vector magnetograms of the central region $ x,y \in
    [-0.5,+0.5] $ for CASE LL1 (left) and CASE LL2 (right). The
    contours represent $B_{z}$. The tangential field is shown by the
    vectors with blue color in positive $B_{z}$ region and red in
    negative $B_{z}$ region.}
  \label{fig:LL_map}
\end{figure}

The NLFFF model derived by \citet{Low1990} has served as standard
benchmark for many extrapolation codes
\citep{Wheatland2000,Amari2006,Schrijver2006,Valori2007,He2008,Jiang2011}.
The fields of this model are basically axially symmetric and can be
represented by a second-order ordinary differential equation derived
in spherical coordinates
\begin{equation}
  \label{eq:low}
  (1-\cos^{2}\theta)\frac{d^{2} P}{ d(\cos
    \theta)^{2}}+n(n+1)P+a^{2}\frac{1+n}{n}P^{1+2/n} = 0
\end{equation}
where $n$ and $a$ are constants. Then the magnetic field is given by
\begin{equation}
  \label{eq:low2}
  B_{r} = \frac{1}{r^{2}\sin\theta}\frac{\partial A}{\partial \theta},
  B_{\theta} = -\frac{1}{r\sin\theta}\frac{\partial A}{\partial r},
  B_{\phi} = \frac{1}{r\sin\theta}Q
\end{equation}
where $A = P(\cos\theta)/r^{n}$ and $Q=aA^{1+1/n}$. The solution $P$
of \Eq~(\ref{eq:low}) is uniquely determined by two eigenvalues, $n$
and its number of nodes $m$ \citep{Low1990,Amari2006}. By arbitrarily
positioning a plane in the space of the analytical fields, one obtains
a different kind of test case, in which the plane represents the
bottom-boundary condition for extrapolation of the overlaying
fields. In this way the fields sliced by the plane show no more
symmetry and thus benefit for a general testing of extrapolation. The
position of the plane is characterized by two additional parameters,
$l$ and $\Phi$. Here we choose two particular solutions characterized
by the parameters ($n,m,l,\Phi$), which are respectively given by
$n=1$, $m=1$, $l=0.3$ and $\Phi=\pi/4$ (referred to as CASE LL1), and
$n=3$, $m=1$, $l=0.3$ and $\Phi=4\pi/5$ (CASE LL2). For both cases,
the computational domain is $ x,y \in [-1,+1] $ and $ z \in [0,2] $
and discretized by uniform grid of $64\times 64\times 64$ (same as
Paper I). The same test solutions are also used by works in the above
references where more analyses of these fields can be found. The
vector magnetograms for both cases at $z=0$ are shown in
\Fig~\ref{fig:LL_map} and their 3D field lines are shown in panels
$(a)$ of \Figs~\ref{fig:LL1_3D} and \ref{fig:LL2_3D}. Basically,
non-potential fields occupy more volume in CASE LL1 than CASE LL2 and
CASE LL2 is `more nonlinear' with a larger $\alpha$ and stronger
fields more concentrated near the center of the model volume. Since
the solutions show rather smooth (small gradients) and relatively
simple magnetic structures with their topologies roughly consistent
with those of the potential fields based on the same LoS magnetograms,
these tests can be regarded as preliminary tests for any new-developed
NLFFF extrapolation methods before facing to more stringent cases or
realistic magnetograms.

\subsection{The van Ballegooijen Reference Model}
\label{sec:ref_AA}

\begin{figure}[\htbp]
  \centering
  \includegraphics[width=\hlenfig]{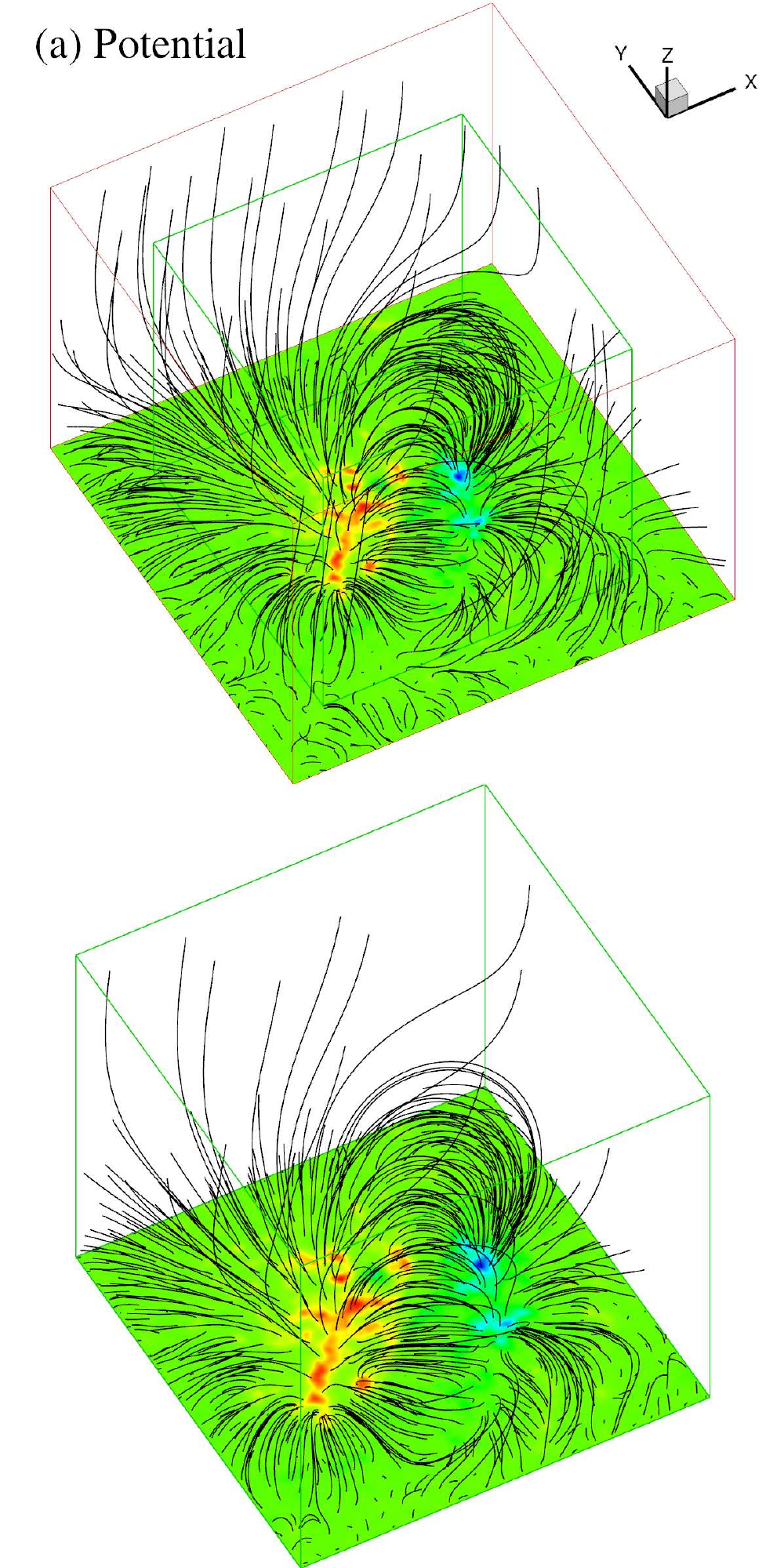}
  \includegraphics[width=\hlenfig]{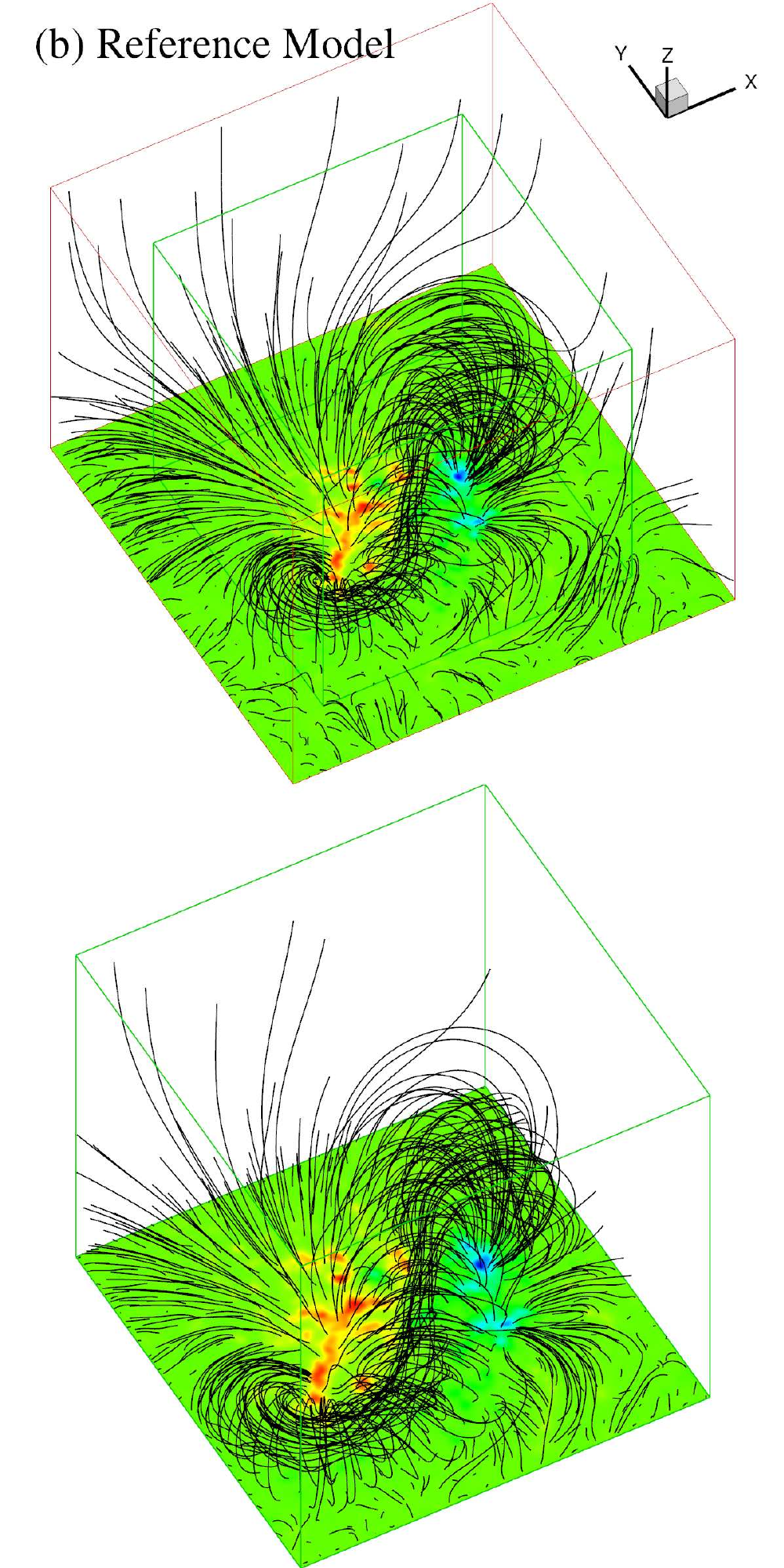}
  \caption{The van Ballegooijen reference model: magnetic field lines
    with contour of $B_{z}$ on the bottom surface. The field lines
    shown are traced from footpoints equally spaced at the bottom
    surface. (a) the initial potential field, (b) the final
    near-force-free reference model. The bottom row enlarges the
    central region outlined by the small cube in the top row.}
  \label{fig:AA}
\end{figure}

The van Ballegooijen reference model is adopted from Paper II for a
more stringent and realistic testing of our code. By this reference
model, it is possible to mimic the analysis of real observational data
while still knowing the properties of the field to be modeled. This
model field is constructed by initially inserting an S-shaped flux
bundle into a potential field associated with active region AR 10814
(see panel (a) of \Fig~\ref{fig:AA}), and then relaxing the unbalanced
system to a near force-free state using van Ballegooijen's
magneto-frictional code in spherical geometry
\citep{Ballegooijen2000,Ballegooijen2004}. Furthermore, an upward
force was applied to the field at the lower boundary during the
relaxation process to mimic the effect of magnetic buoyancy in the
photoshpere, and thus achieve more realistic magnetic fields between
photospheric and chromospheric heights in the model. Finally, by
coordinate transformation and interpolation from the original spherical
geometry, magnetic fields in a Cartesian box of $320\times 320\times
258$ pixels centered on the active region is extracted as the final
reference model. The final near-force-free field are drawn in panel
(b) of \Fig~\ref{fig:AA}, which contains several interesting
topological features including a coronal null and its associated
separatrix surface, and the S-shaped flux bundle surrounded by a
quasi-separatrix layer (please see more details in Paper II). Compared
with the initial potential field, the magnetic topology near the
bottom is significantly modified by the low-lying flux rope, which
challenges the extrapolation much more than the Low and Lou
cases. Because of extra force presented at the bottom, this reference
model can be used for tests of extrapolations from either the
`chromospheric' or the `photospheric' magnetograms, by providing the
NLFFF code with data at $z=z_{2}$ or $z=z_{0}$ ($z$ is the height in
the model, e.g., $z_{0}$ is the base of the reference model). For the
`chromospheric' case, the boundary data used is largely force-free
which is consistent with the extrapolation method, while the
`photospheric' case is more forced and thus represents a more
realistic magnetogram of observation. It should be stressed that the
van Ballegooijen reference model is not strictly force-free in the
whole model box, even above the chromosphere, due to the
implementation of the magneto-frictional method and some other
numerical errors. It is demonstrated that in the model the residual
forces of at least $5\%$ of magnetic-pressure force are present up to
height of $z_{30}$, which are consistent with what is known of forces
on the Sun, making the model a realistic, solar-like test case for the
extrapolation codes.

\begin{figure}[\htbp]
  \centering
  \includegraphics[width=\textwidth]{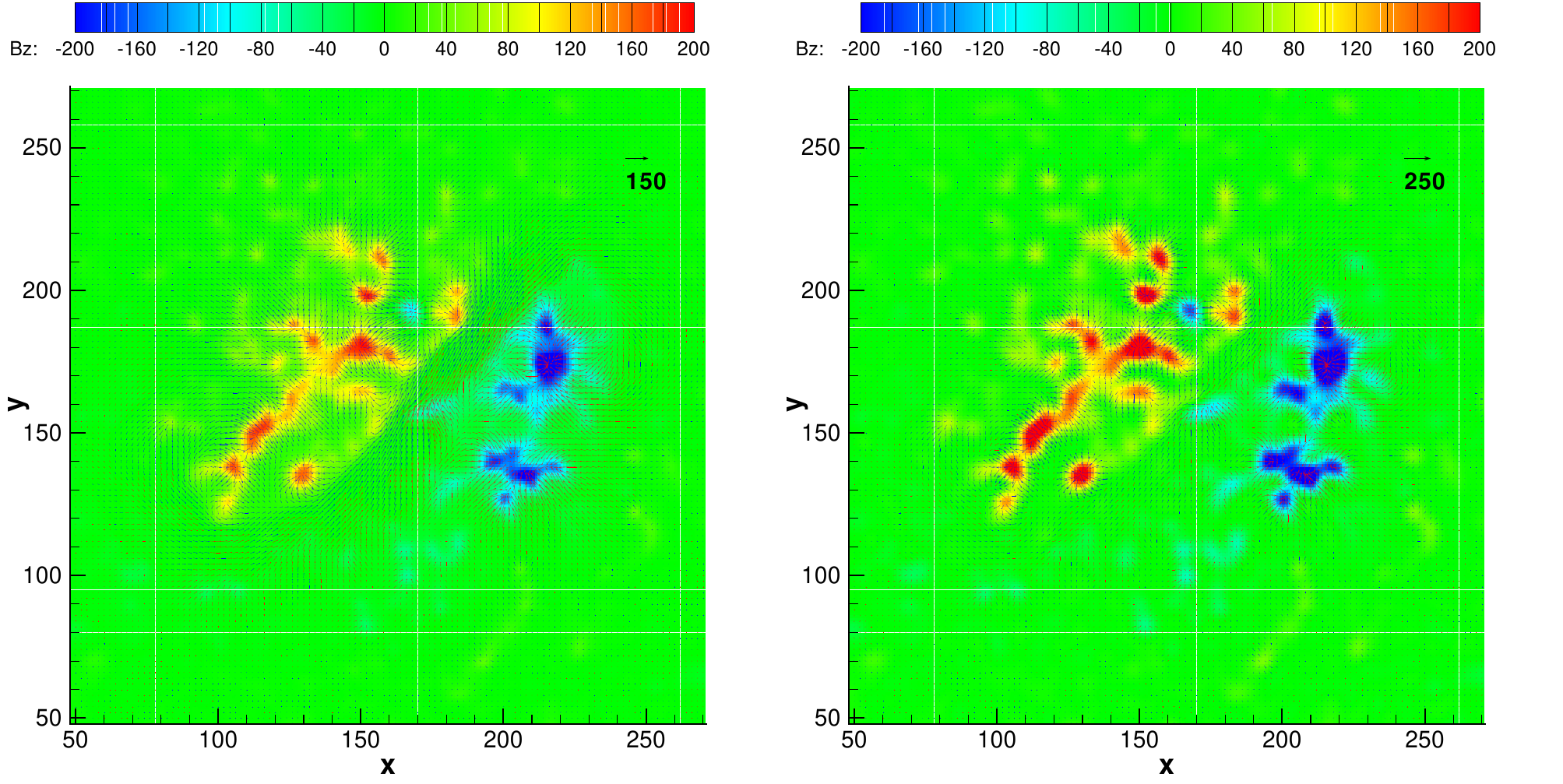}
  \caption{Vector magnetograms of the central $224^{2}$ pixels for
    chromospheric case (left) and preprocessed photospheric case
    (right). The contours represent $B_{z}$ with a saturation level of
    $\pm 200$ G. The tangential field is shown by the vectors (plotted
    at every second grid point) with blue color in positive $B_{z}$
    region and red in negative $B_{z}$ region.}
  \label{fig:AA_map}
\end{figure}

As done in Paper II, we will test our code by both the chromospheric
and photospheric cases. For the photospheric case which is
inconsistent with force-free assumption, we only examine the code with
the photospheric magnetogram preprocessed by method of
\citet{Wiegelmann2006}. The magnetograms of both test cases are
plotted in \Fig~\ref{fig:AA_map}, which show a significant shearing
along the polarity inversion line (PIL).

\section{Metrics}
\label{sec:metrics}

For a detailed analysis of the extrapolation fields, a suite of
metrics introduced in \citep{Schrijver2006} are computed. These
metrics compare either local characteristics including vector
magnitudes and directions at each point or the global energy
content. They are respectively the vector correlation $C_{\rm vec}$
\begin{equation}
  \label{eq:test.1}
  C_{\rm vec} \equiv
  \sum_{i}\mathbf{B}_{i}\cdot\mathbf{b}_{i}/
  (\sum_{i}|\mathbf{B}_{i}|^{2}\sum_{i}|\mathbf{b}_{i}|^{2}),
\end{equation}
the metric $C_{\rm CS}$ based on the Cauchy-Schwarz inequality
\begin{equation}
  \label{eq:test.2}
  C_{\rm CS} \equiv
  \frac{1}{M}\sum_{i}\frac{\mathbf{B}_{i}\cdot\mathbf{b}_{i}}
  {|\mathbf{B}_{i}||\mathbf{b}_{i}|},
\end{equation}
the normalized and mean vector error $E_{\rm n}'$, $E_{\rm m}'$
\begin{eqnarray}
  \label{eq:test.3}
  E_{\rm n} \equiv
  \sum_{i}|\mathbf{b}_{i}-\mathbf{B}_{i}|/\sum_{i}|\mathbf{B}_{i}|;
  E_{\rm n}' = 1-E_{\rm n}, \\
  E_{\rm m} \equiv
  \frac{1}{M}\sum_{i}\frac{|\mathbf{B}_{i}-\mathbf{b}_{i}|}
  {|\mathbf{B}_{i}|}; E_{\rm m}' = 1-E_{\rm m},
\end{eqnarray}
where $\mathbf{B}_{i}$ and $\mathbf{b}_{i}$ denote the input field
(the Low \& Lou's solution or the van Ballegooijen reference model in
this paper) and the extrapolated field, respectively, $i$ denotes the
indices of the grid points and $M$ is the total number of grid points
involved. As can be seen, an exact extrapolation will have all the
metrics equal to unity in such definitions, and the closer to unity
means the better extrapolation and vice versa. Detailed descriptions
for these metrics can be found in
\citep{Amari2006,Schrijver2006,Valori2007}. Another very important
parameter for comparing the extrapolation is the free energy of
magnetic field. It is measured by the ratio of extrapolated energy to
the potential energy using the same magnetogram,
\begin{equation}
  \label{eq:test.4}
  E/E_{\rm pot} = \frac{\sum_{i}|\vec B_{i}|^{2}}
  {\sum_{i}|(\vec B_{\rm pot})_{i}|^{2}},
\end{equation}

It is common to measure the force-freeness and divergence-freeness of
the extrapolation using the current-weighted sine metric CWsin and
divergence metric $\langle |f_{i}|\rangle$
\citep{Metcalf2008,Schrijver2008,Derosa2009,Canou2010}, which are
defined by \citet{Wheatland2000} as
\begin{equation}
  {\rm CWsin} \equiv
  \frac{\sum_{i}|\mathbf{J}_{i}|\sigma_{i}}{\sum_{i}|\mathbf{J}_{i}|};
  \sigma_{i} =
  \frac{|\mathbf{J}_{i}\times\mathbf{B}_{i}|}
  {|\mathbf{J}_{i}||\mathbf{B}_{i}|}
\end{equation}
and 
\begin{equation}
  \langle |f_{i}|\rangle =
  \frac{1}{M}\sum_{i}\frac{(\divB)_{i}}{6|\vec B_{i}|/\Delta x}
\end{equation}
where $\Delta x$ is the grid spacing. Both of the metrics are
normalized with the former focused on the directional deviation
between the currents and the field lines and the latter on the
relative value of residual divergence. These two metrics are equal to
zero for an exact force-free field, and hence the smaller these
metrics are, the better the extrapolation is.

In addition to the above metrics, we also introduce another pair of
metrics to evaluate the degree of convergence towards the
divergence-free and force-free state. For the first one, we note that
a nonzero $\divB$ (i.e., the magnetic monopole) introduces to the
system an unphysical force $\vec F=\vec B\divB$ parallel to the field
line \citep{Dellar2001}. To evaluate the effect of this unphysical
force to the numerical computation, the metric $E_{\divB}$ is defined
as the average ratio of this force to the magnetic-pressure force
\begin{equation}
  E_{\divB} = \frac{1}{M}\sum_{i}\frac{|\mathbf{B}_{i}(\divB)_{i}|}
  {|\nabla(|\vec B|^{2}/2)_{i}|}.
\end{equation}
Similarly, the second metric $E_{\crlB}$ measures the effect of the
residual Lorentz force in the same way
\begin{equation}
  E_{\crlB} = \frac{1}{M}\sum_{i}\frac{|\vec J_{i}\times\vec B_{i}|}
  {|\nabla(|\vec B|^{2}/2)_{i}|}.
\end{equation}
Unlike the metrics of CWsin and $\langle |f_{i}|\rangle$ which mainly
characterize the geometric properties of the field, these two metrics
directly measure the physical action of the residual divergence and
Lorentz force on the system in the actual numerical computation. This
is important for checking of the NLFFF solution if it is used to
initialize any MHD simulations.

For the all metrics above, the first four are more rigid since they
are involved without any type of derivatives, while the other metrics
may be unreliable for comparison with results from different papers
due to the specific numerical realization of the derivatives (for
example, different orders of numerical differentiation or different
configurations of computational grid, e.g., cell-centered or
staggered). In the present work, the second-order central difference
is used for evaluating all the derivatives associated with the
divergence, curl, and gradient operators, although the spatial
derivatives can be directly obtained from the CESE method.

\section{Results}
\label{sec:results}

In this section, we present the results of extrapolation for the
benchmark models. The results are also compared with some results
reported in Paper I, II and \citep{Valori2007}.

\subsection{Low and Lou's Force Free Field}
\label{sec:low}

\subsubsection{CASE LL1}

\begin{figure}[\htbp]
  \centering
  \includegraphics[width=\textwidth]{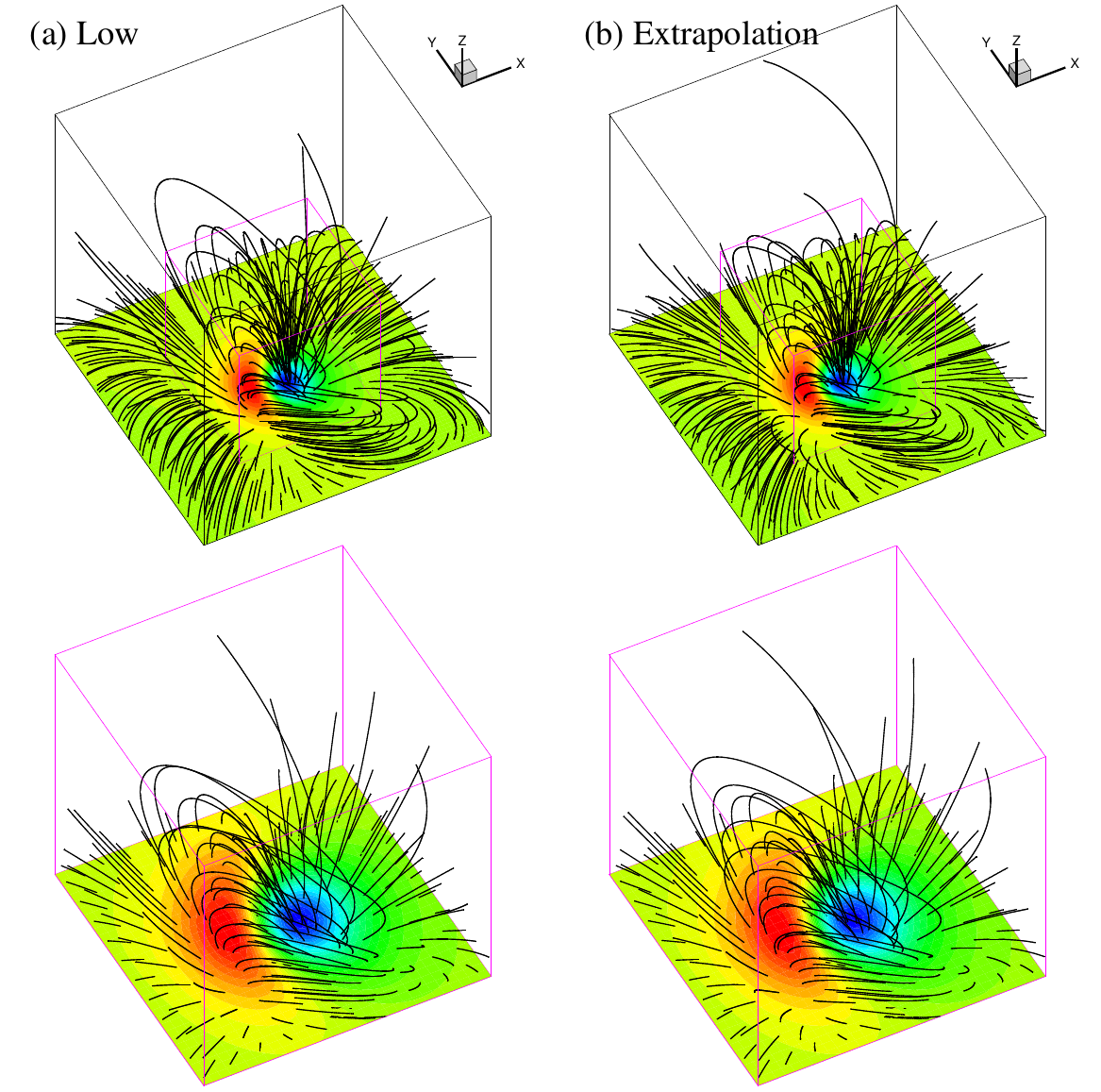}
  \caption{CASE LL1: magnetic field lines with contour of $B_{z}$ on
    the bottom surface. (a) the Low \& Lou's solution, (b) the
    extrapolation result. The bottom row enlarges the central region
    ($ x,y \in [-0.5,+0.5] $ and $ z \in [0,1] $) outlined by the
    small cube in the top row.}
  \label{fig:LL1_3D}
\end{figure}

\begin{figure}[\htbp]
  \centering
  \includegraphics[width=\textwidth]{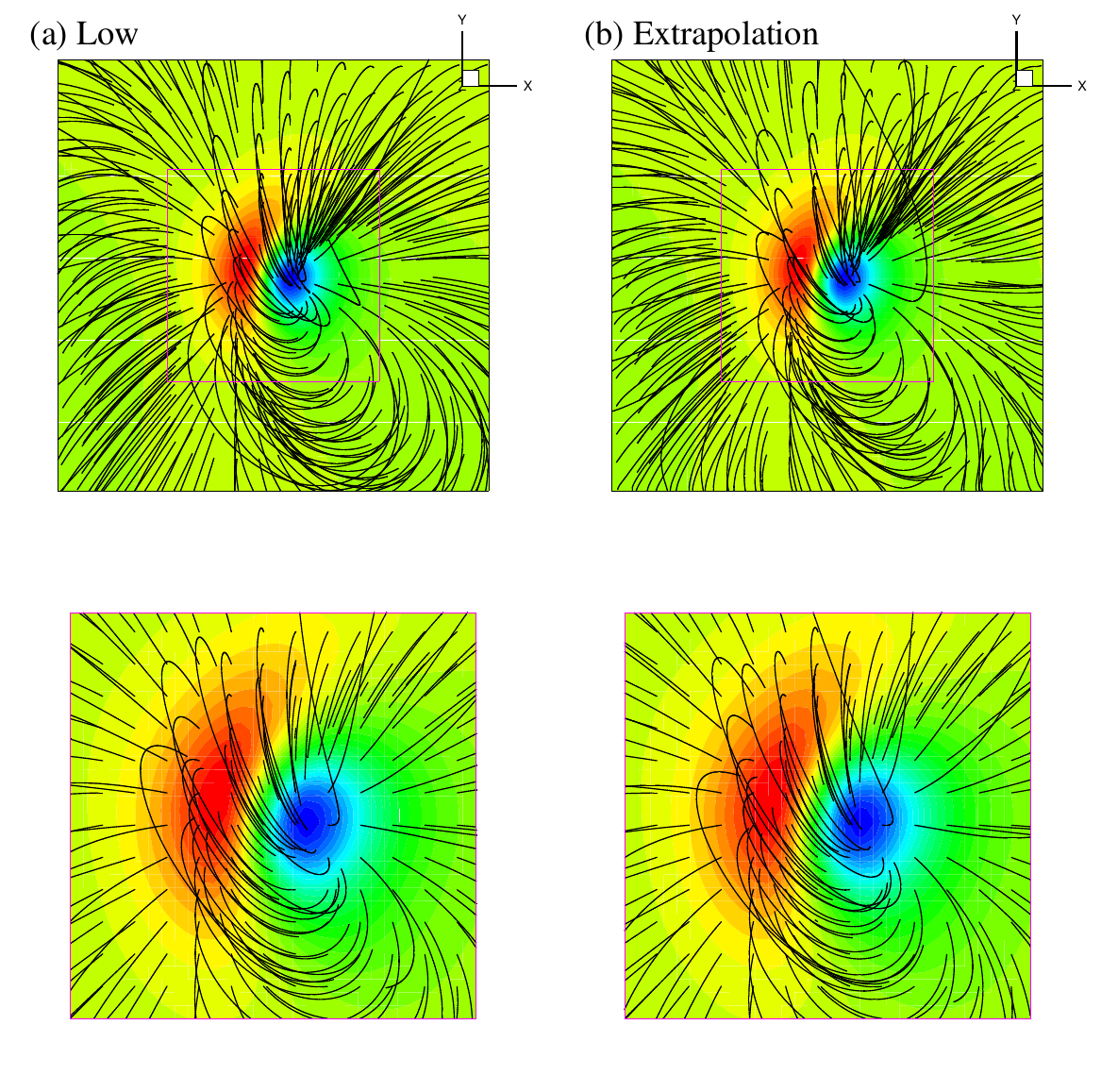}
  \caption{CASE LL1: same as \Fig~\ref{fig:LL1_3D}, but projected onto
    the $x$--$y$ planes.}
  \label{fig:LL1_xy}
\end{figure}

Results for CASE LL1 are given in \Fig~\ref{fig:LL1_3D},
\Fig~\ref{fig:LL1_xy}, Table~\ref{tab:LL1} and
Table~\ref{tab:LL1_divB}. In \Fig~\ref{fig:LL1_3D}, we show the same
selected field lines in 3D view for the extrapolation results and the
reference solution, which are traced from foot-points evenly rooted at
the lower boundary. In the central region of core fields (i.e., $ x,y
\in [-0.5,+0.5] $ and $ z \in [0,1] $, enlarged in the bottom row of
the figures), the MHD result is highly in agreement with Low \& Lou's
solution, as can be seen from the high similarity of most of the field
lines. Such agreement is demonstrated quantitatively by the metrics in
Table~\ref{tab:LL1}. The first three metrics are very close to $1$
with error of $<5\%$ and even the most sensitive metric $E_{m}'$ has
an error below $10\%$. In Table~\ref{tab:LL1} we also compare with the
best result by Wiegelmann's code \citep{Wiegelmann2004} reported in
Paper I and extrapolation by \citet{Valori2007}. Our result for the
central region, although only specified the lower boundary, still
reaches the level of the best extrapolation that using information of
Low \& Lou's solution on all six boundaries. This may be due to the
fact that this test case is close to potential field hence a fixed
side and top boundary conditions can rarely impact the central
extrapolation. The influence of the boundary conditions is more
explicitly shown by comparison of the metrics for the entire
domain. Now the Wiegelmann's extrapolation still performs perfectly
with all four metrics extremely close to unity, while our result and
the Valori's perform less exactly, but still satisfactory. Like other
results in the Table, our extrapolation also recovered the energy
content very precisely, especially for the central region; furthermore
the metrics evaluating the force-freeness and divergence-freeness are
rather small and close to the reference values which is caused by
discretization error. All these shows that extrapolation of very high
accuracy can be achieved by our implementation, at least for the
present, relatively easy test case.

In the Valori's implementation of the magneto-frictional method
\citep{Valori2007}, a fourth-order numerical scheme is used with many
layers of ghost mesh and a high-order polynomial extrapolation of the
fields is adopted on the side and top boundaries. However by comparing
our result with Valori's, it is interesting to note that our
implementation performs better although our numerical scheme is
second-order method without any ghost layers and the boundaries are
simply fixed. In this test, the boundary effect can be neglected as
said (while in the following test of CASE LL2, we will see the effect
of different treatments of these boundaries). Then we concluded that
the better performance is due to the merit of using a better solver,
i.e., CESE method combined with the magnetic splitting algorithm which
can gain additional accuracy.

\begin{figure}[\htbp]
  \centering
  \includegraphics[width=\textwidth]{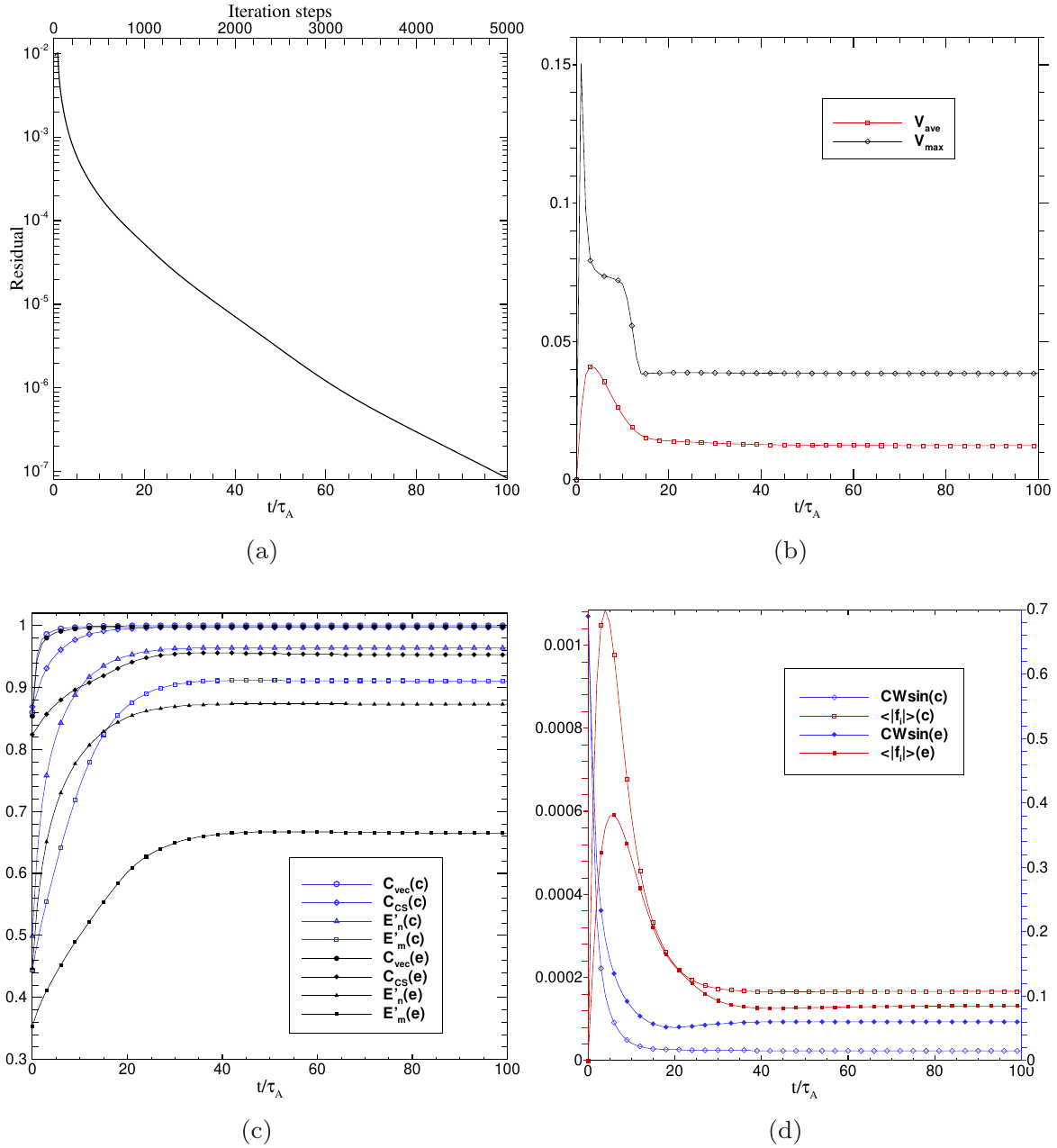}
  \caption{CASE LL1: The history of the relaxation to force-free
    equilibrium. (a) Evolution of residual ${\rm res}(\vec B)$ with
    time (or the iteration steps); (b) evolution of the maximum and
    average velocity; and (c), (d) evolution of the metrics for the
    central region (marked by (c)) and the entire volume (marked by
    (e)).}
  \label{fig:LL1_Convergence}
\end{figure}

We finally give a study of the convergence of the extrapolation.  In
\Fig~\ref{fig:LL1_Convergence} we shows the history of the system
relaxing to the final force-free equilibrium, including the residual
of temporal evolution of the magnetic field
\begin{equation}
  \label{eq:res}
  {\rm res}^{n}(\vec B) = \sqrt{\frac{1}{3}\sum_{\delta = x,y,z}
    \frac{\sum_{i}(B_{i\delta}^{n}-B_{i\delta}^{n-1})^{2}}{\sum_{i}(B_{i\delta}^{n})^{2}}}
\end{equation}
(where $n$ denotes the iteration steps), the evolution of the velocity
and the metrics. The system converged very fast from a initial
residual of $10^{-2}$ to value under $10^{-7}$ with time of
$100\tau_{A}$ (about $5000$ iteration steps, see panel (a) of
\Fig~\ref{fig:LL1_Convergence}).
The evolution of the plasma velocity indicates that a static
equilibrium is reached as expected with a rather small residual
velocity $\sim 0.01$ which is only on the order of the numerical error
$O(\Delta x^{2})$ of the CESE solver. All the metrics plotted in the
figure converged after $40\tau_{A}$ (about 2000 iteration steps), when
the residual is on the order of $10^{-5}$. Note that the metric
$f_{i}$ of $\divB$, like the plasma velocity, first climbs to a
relatively high level ($\sim 0.01$, see panel (d)) and then drops to
the level of discretization errors. In principle the divergence-free
constraint of $\vec B$ should be fulfilled throughout the evolution,
at least close to level of discretization error. However, a ideally
dissipationless induction equation \Eq~(\ref{eq:m.4}) with
divergence-free constraint can preserve the magnetic connectivity,
which makes the topology of the magnetic field unchangeable
\citep{Wiegelmann2008} unless a finite resistivity is included for
allowing reconnection and changing of the magnetic topology
\citep{Roumeliotis1996}. In the present implementation in which no
resistivity is included in the induction equation, a break of $\divB$
constraint in the initial evolution process (indicated by the climb of
metric $f_{i}$) can thus produce a change in the magnetic topology
(also note that a numerical diffusion can help to topology adjustment
in addition).

\begin{table}[htbp]
  \centering
  \begin{tabular}{llllllll}
    \hline
    \hline
    Model & $C_{\rm vec}$ & $C_{\rm CS}$ & $E_{\rm n}'$ & $E_{\rm m}'$
    & $E/E_{\rm pot}$ & CWsin & $\langle |f_{i}|\rangle$ \\
    \hline
    For the central region \\
    Low                 & 1.000 & 1.000 & 1.000 & 1.000 & 1.242 & 0.014 & $0.94\times 10^{-4}$\\
    {\bf Our result}    & 1.000 & 0.997 & 0.964 & 0.912 & 1.241 & 0.015 & $1.67\times 10^{-4}$ \\
    Wiegelmann$^{*}$     & 1.00  & 1.00  & 0.97  & 0.96  & 1.26  & & \\
    {\bf Valori}$^{**}$  & 0.999 & 0.99  & 0.95  & 0.87  & 1.23 & 0.009 &\\
    Potential           & 0.858 & 0.869 & 0.498 & 0.443 & 1.000 &  &\\ 
    \hline
    For the entire domain \\
    Low                 & 1.000 & 1.000 & 1.000 & 1.000 & 1.294 & 0.014 & $0.56\times 10^{-4}$\\
    {\bf Our result}    & 0.998 & 0.955 & 0.873 & 0.662 & 1.282 & 0.060 & $1.31\times 10^{-4}$\\
    Wiegelmann$^{*}$     & 1.00  &  1.00 & 0.98  & 0.98 & 1.31   & 0.070  &\\
    {\bf Valori$^{**}$}  & 0.994 & 0.86  & 0.80 & 0.51   & 1.28  & 0.019 &\\
    Potential           & 0.852 & 0.824 & 0.446 & 0.353 & 1.000 & &\\
    \hline
  \end{tabular}
  \caption{CASE LL1: Metrics for the central and entire regions. The
    superscript $*$ denotes the reported results in Table I and Table
    II of Paper I, and $**$ denotes the reported results by
    \citet{Valori2007}.}
  \label{tab:LL1}
\end{table}

\begin{table}[htbp]
  \centering
  \begin{tabular}{lll}
    \hline
    \hline
    Model & $E_{\crlB}$ & $E_{\divB}$ \\
    \hline
    For the central region \\
    Low & 0.007 & 0.005\\
    Our result& 0.010 & 0.008\\
    \hline
    For the entire region \\
    Low & 0.003 & 0.002\\
    Our result& 0.024 & 0.005\\
    \hline
  \end{tabular}
  \caption{CASE LL1: Metrics of $E_{\crlB}$ and $E_{\divB}$}
  \label{tab:LL1_divB}
\end{table}

\subsubsection{CASE LL2}

\begin{figure}[\htbp]
  \centering
  \includegraphics[width=\textwidth]{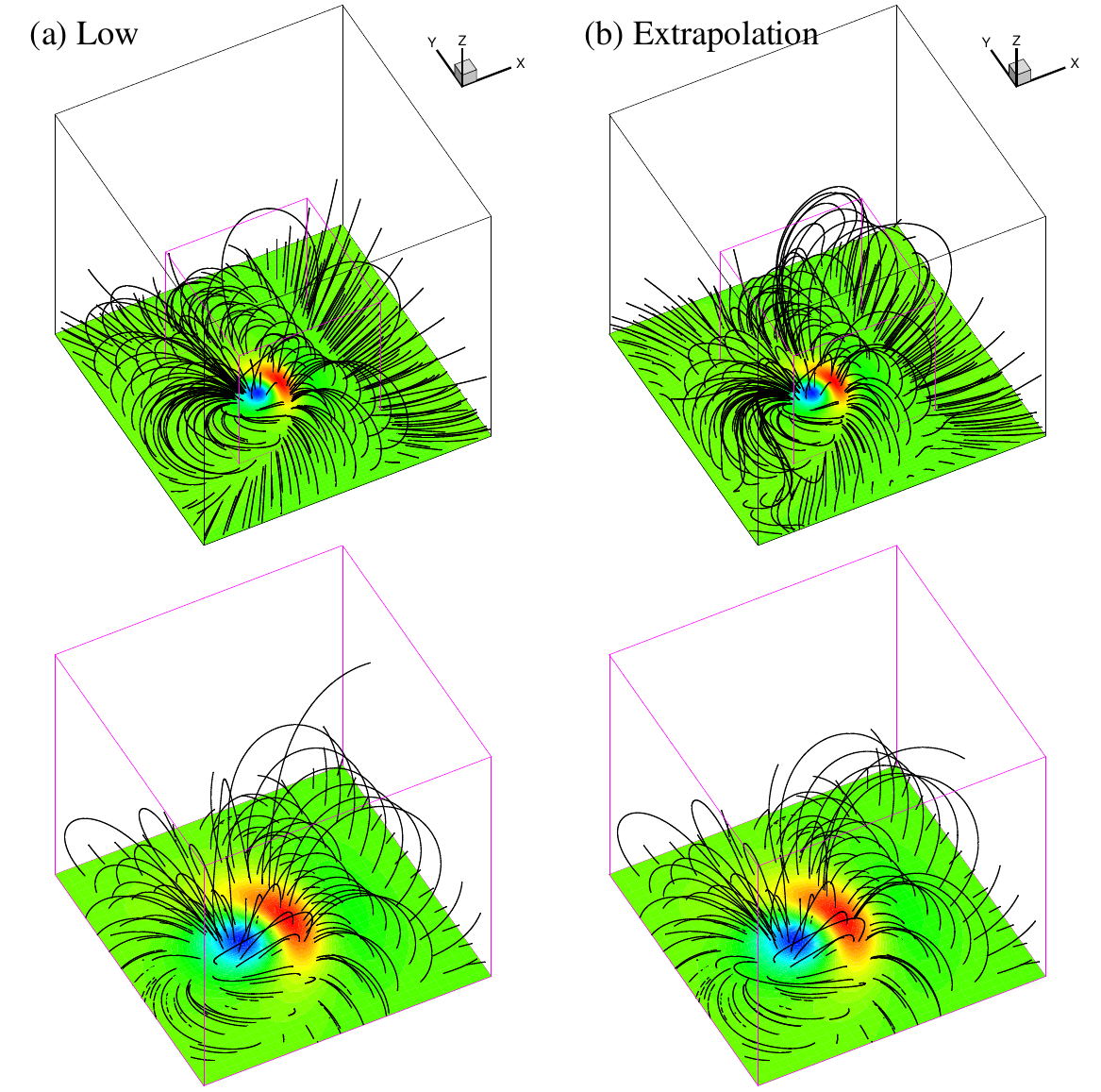}
  \caption{Same as \Fig~\ref{fig:LL1_3D} but for CASE LL2.}
  \label{fig:LL2_3D}
\end{figure}

\begin{figure}[\htbp]
  \centering
  \includegraphics[width=\textwidth]{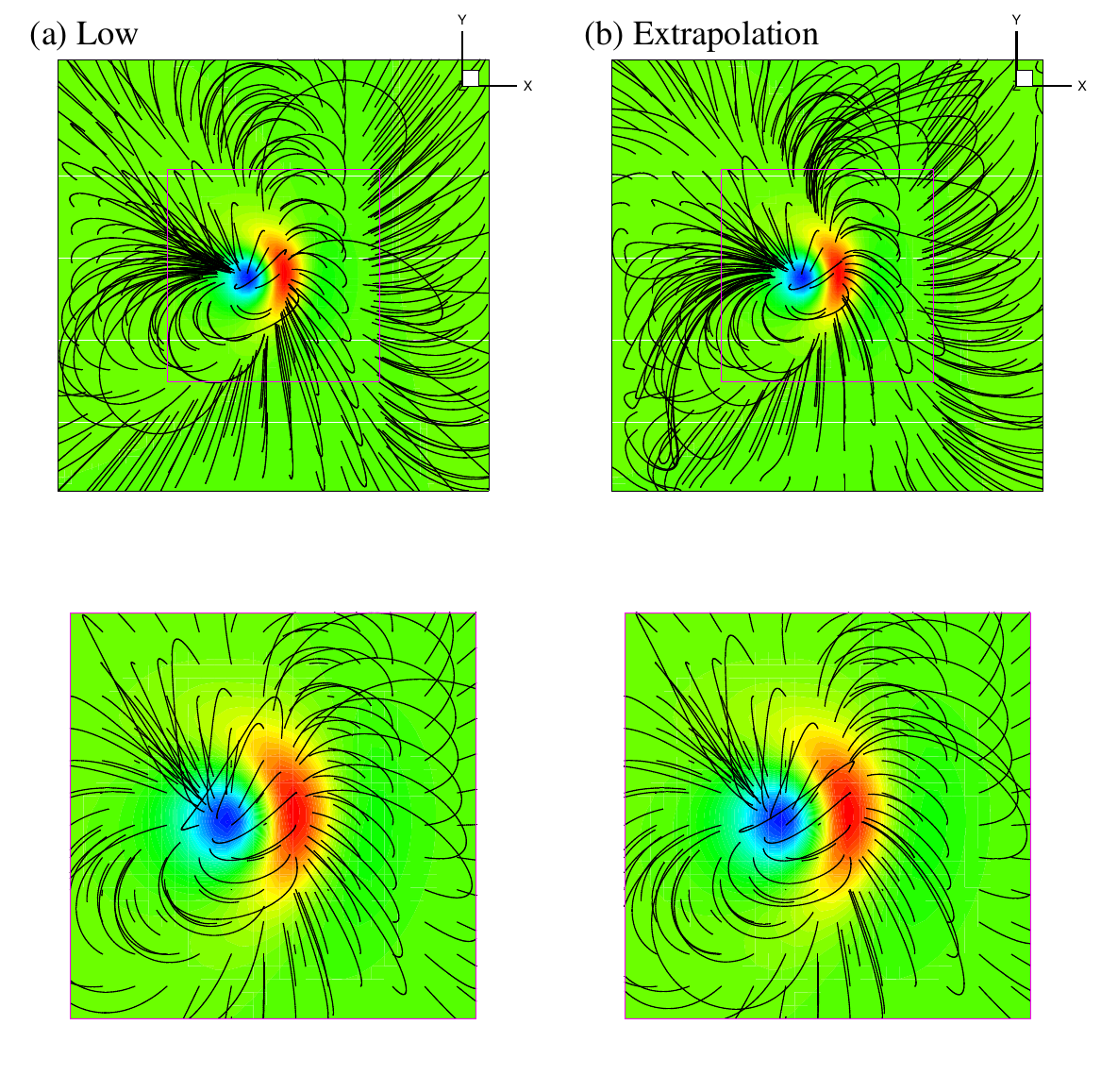}
  \caption{Same as \Fig~\ref{fig:LL1_xy} but for CASE LL2.}
  \label{fig:LL2_xy}
\end{figure}

Now we present the result of the second test CASE LL2 which is more
difficult than the first one. Compared with CASE LL1, this case is more
non-potential and nonlinear with relatively larger gradient of
fields. Although considering of this, our extrapolation still gives
satisfactory result that is as good as the best result in Paper I (see
the first four metrics shown and compared in Table~\ref{tab:LL2}). It
is also quite encouraging that our result for the central region is
close to Valori's which is computed with fourth-order numerical
scheme. \Fig~\ref{fig:LL2_3D} and \ref{fig:LL2_xy} demonstrate
qualitatively the consistence with the Low and Lou's reference
solution. The energy contents are very well reproduced for both the
central region and the entire domain.

By comparing the metrics of the entire domain, we find that our result
scores worse than the Valori's, especially shown by $E'_{\rm m}$. The
reason for this is twofold. Firstly, a high-order scheme can
characterize the large gradient much more accurately than 2nd-order
scheme and thus the fourth-order scheme by \citet{Valori2007} shows
its advantages for this test case with larger gradient than CASE LL1
(the high-order accuracy can be also achieved by mesh refinement other
than improving the order of the numerical scheme. As demonstrated in
our previous work \citep{Jiang2011}, a refined grid of $128\times
128\times 100$ with the CESE scheme gave much more accurate result
with the most sensitive metric even reaching $0.166$). Secondly, our
implementation simply fixed the artificial boundaries (i.e., the
lateral and top faces), which obviously make the system
over-determined. In the present test case the final field is very
non-potential, hence the boundary values deviate from the initial
conditions significantly. A fixed boundary condition will tie the
field lines that pass through the boundary, which would otherwise
freely cross the boundary since this boundary is a non-existing
interface in the realistic corona. This line-tied condition thus
hinders the system from relaxing to a true force-free state. The
metric of CWsin demonstrates clearly that the field is less force-free
than the Valori's result. The boundary effect can be seen visually
from some field lines close to the lateral boundaries shown in the
figures (e.g., see the distortion of field lines at the lower left
coroners of panel (b) in \Fig~\ref{fig:LL2_3D} and
\ref{fig:LL2_xy}). Also caused by the boundary effect, the field for
the entire domain (${\rm CWsin}=0.060$) is thus less force-free than
the central region (${\rm CWsin}=0.047$). In a full MHD simulation,
this boundary effect can be minimized by using a so-called
non-reflecting boundary conditions based on characteristic
decomposition of the full MHD system
\citep{Wu2001,Wu2006,Hayashi2005,Feng2010,Jiang2011}. However for the
present form of magneto-frictional equation, the characteristic method
is no more valid because of the eigen degeneration of the
non-hyperbolic system. In such situation, a natural choice of modeling
the non-reflecting boundary is to use linear or high-order
extrapolation, just as done by \citet{Valori2007}. Fortunately, field
configuration like this Low \& Lou case with the entire volume very
non-potential is not usually found in observed magnetograms, and by
choosing a model box significant larger than the core non-potential
region, a simply fixed boundary values of the initial potential field
is still sufficient for most extrapolations, for example, test cases
in the next section.

\begin{figure}[\htbp]
  \centering
  \includegraphics[width=\textwidth]{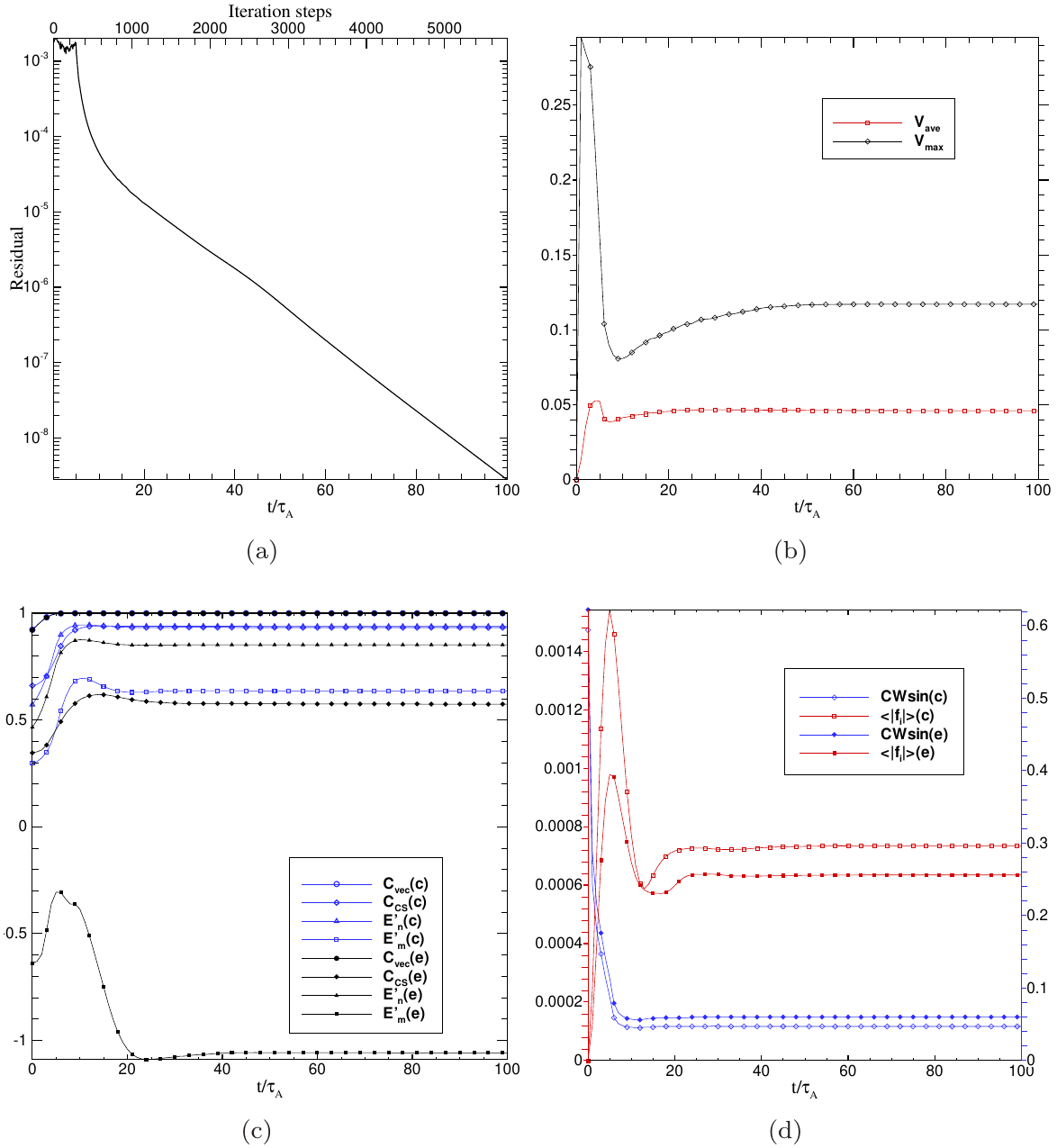}
  \caption{Same as \Fig~\ref{fig:LL1_Convergence} but for CASE LL2.}
  \label{fig:LL2_Convergence}
\end{figure}

\Fig~\ref{fig:LL2_Convergence} shows that the convergence speed is
even faster than case LL1. The residual of the magnetic field reaches
the order of $10^{-7}$ with only $\sim 3200$ iterations at
$60\tau_{A}$. The evolution speed of the magnetic field is also
demonstrated by magnitude of the plasma velocity, which is larger than
that of case LL1. At about $30\tau_{A}$, all the metrics converged
with the corresponding residual of magnetic field on the order of
$10^{-5}$. This and the preceding cases, shows that generally the
iteration can be stopped when ${\rm res}^{n}(\vec B)< 10^{-6}$ since
after that temporal change of the magnetic field can be neglected
actually.

\begin{table}[htbp]
  \centering
  \begin{tabular}{llllllll}
    \hline
    \hline
    Model & $C_{\rm vec}$ & $C_{\rm CS}$ & $E_{\rm n}'$ & $E_{\rm m}'$
    & $E/E_{\rm pot}$ & CWsin & $\langle |f_{i}|\rangle$ \\
    \hline
    For the central region \\
    Low                 & 1.000 & 1.000 & 1.000 & 1.000 & 1.099 & 0.036 & $3.14\times 10^{-4}$\\
    {\bf Our result}    & 0.999 & 0.933 & 0.938 & 0.636 & 1.114 & 0.047 & $7.35\times 10^{-4}$\\
    Wiegelmann$^{*}$     & 1.00  & 0.91  & 0.92  & 0.66  & 1.14 & & \\
    {\bf Valori}$^{**}$ & 0.999 & 0.95  & 0.96  & 0.75  & 1.11 & 0.015 &\\
    Potential          & 0.923 & 0.661 & 0.572 & 0.299 & 1.000 &  &\\ 
    \hline
    For the entire domain \\
    Low                 & 1.000 & 1.000 & 1.000 & 1.000 & 1.100 & 0.035 & $2.17\times 10^{-4}$\\
    {\bf Our result}    & 0.999 & 0.574 & 0.852 & -1.060 &1.115 & 0.060 & $6.35\times 10^{-4}$\\
    Wiegelmann$^{*}$     & 1.00  &  0.57 & 0.86  & -0.25 & 1.14 &  &\\
    {\bf Valori$^{**}$}  & 0.999 & 0.64  & 0.88 & -0.01   & 1.11 & 0.027&\\
    Potential           & 0.921 & 0.346 & 0.465 &-0.639  &1.000 & &\\
    \hline
  \end{tabular}
  \caption{Same as Table~\ref{tab:LL1} but for LL CASE2.  The
    superscript $*$ denotes the reported results in Table I and Table
    II of Paper I, and $**$ denotes the reported results by
    \citet{Valori2007}.}
  \label{tab:LL2}
\end{table}

\begin{table}[htbp]
  \centering
  \begin{tabular}{lll}
    \hline
    \hline
    Model & $E_{\crlB}$ & $E_{\divB}$ \\
    \hline
    For the central region \\
    Low       & 0.011 & 0.008\\
    Our result& 0.026 & 0.024\\
    \hline
    For the entire region \\
    Low       & 0.004 & 0.003\\
    Our result& 0.025 & 0.013\\
    \hline
  \end{tabular}
  \caption{CASE LL2: Metrics of $E_{\crlB}$ and $E_{\divB}$}
  \label{tab:LL2_divB}
\end{table}

\subsection{The van Ballegooijen Reference Field}
\label{sec:AA}

\begin{figure}[\htbp]
  \centering
  \includegraphics[width=\textwidth]{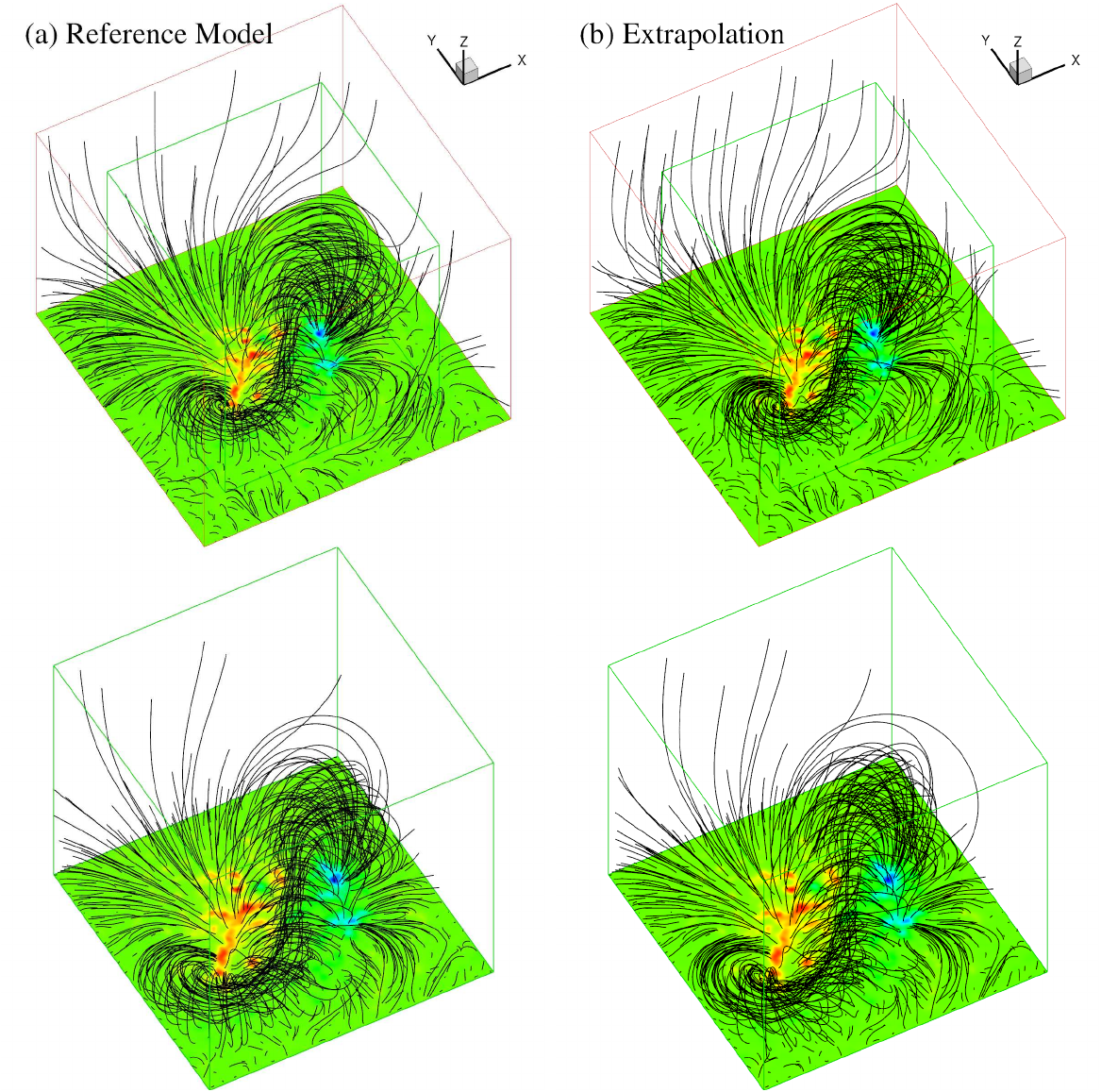}
  \caption{Chromospheric test case of the van Ballegooijen reference
    model: magnetic field lines with contour of $B_{z}$ on the bottom
    surface. (a) the reference model, (b) the extrapolation
    result. The bottom row enlarges the central region outlined by the
    small cube in the top row.}
  \label{fig:AA_chrom_3D}
\end{figure}

\begin{figure}[\htbp]
  \centering
  \includegraphics[width=\textwidth]{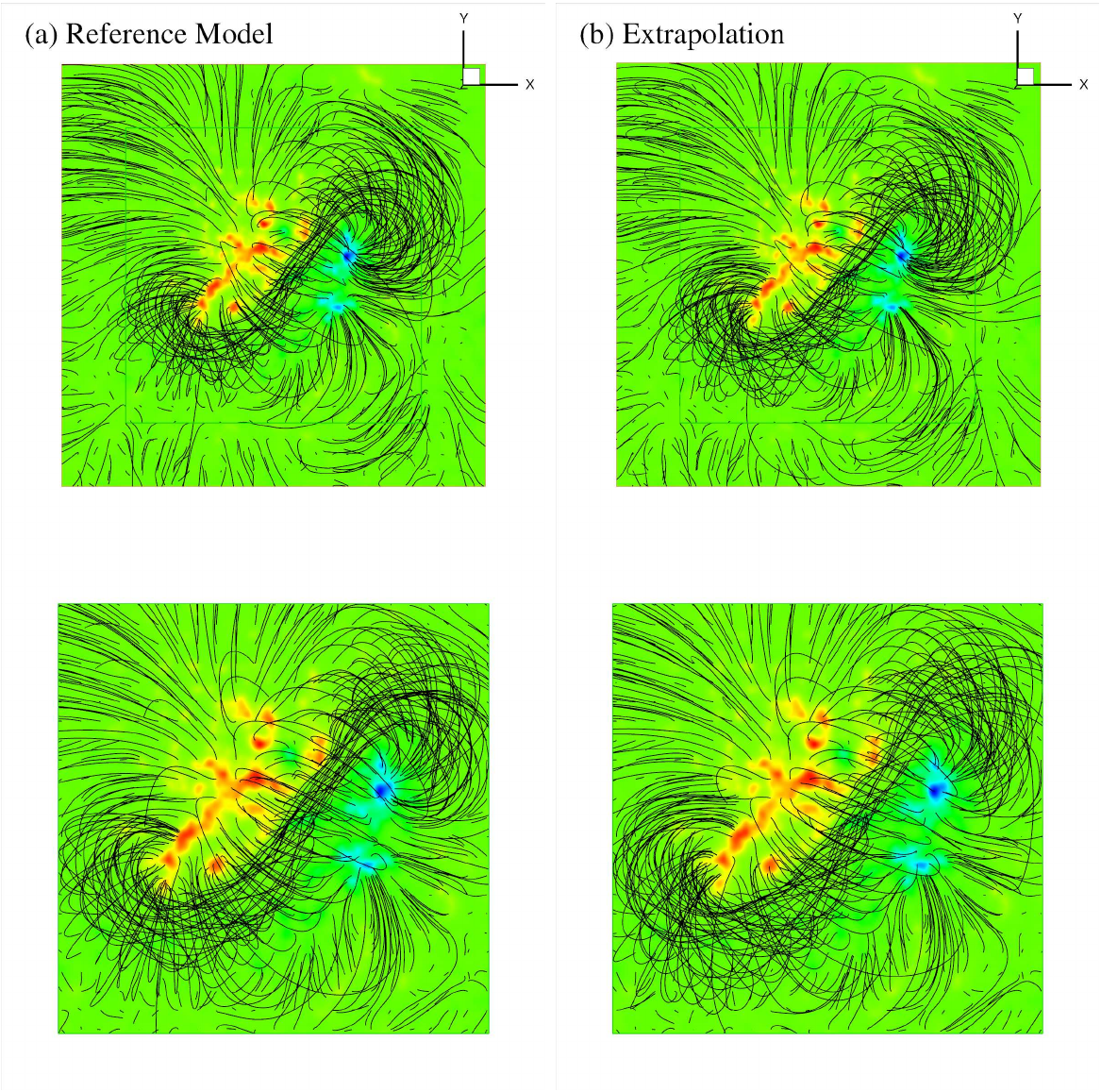}
  \caption{Chromospheric test case of the van Ballegooijen reference
    model: same as \Fig~\ref{fig:AA_chrom_3D}, but projected onto the
    $x$--$y$ planes.}
  \label{fig:AA_chrom_xy}
\end{figure}

\begin{figure}[\htbp]
  \centering
  \includegraphics[angle=270,width=\textwidth]{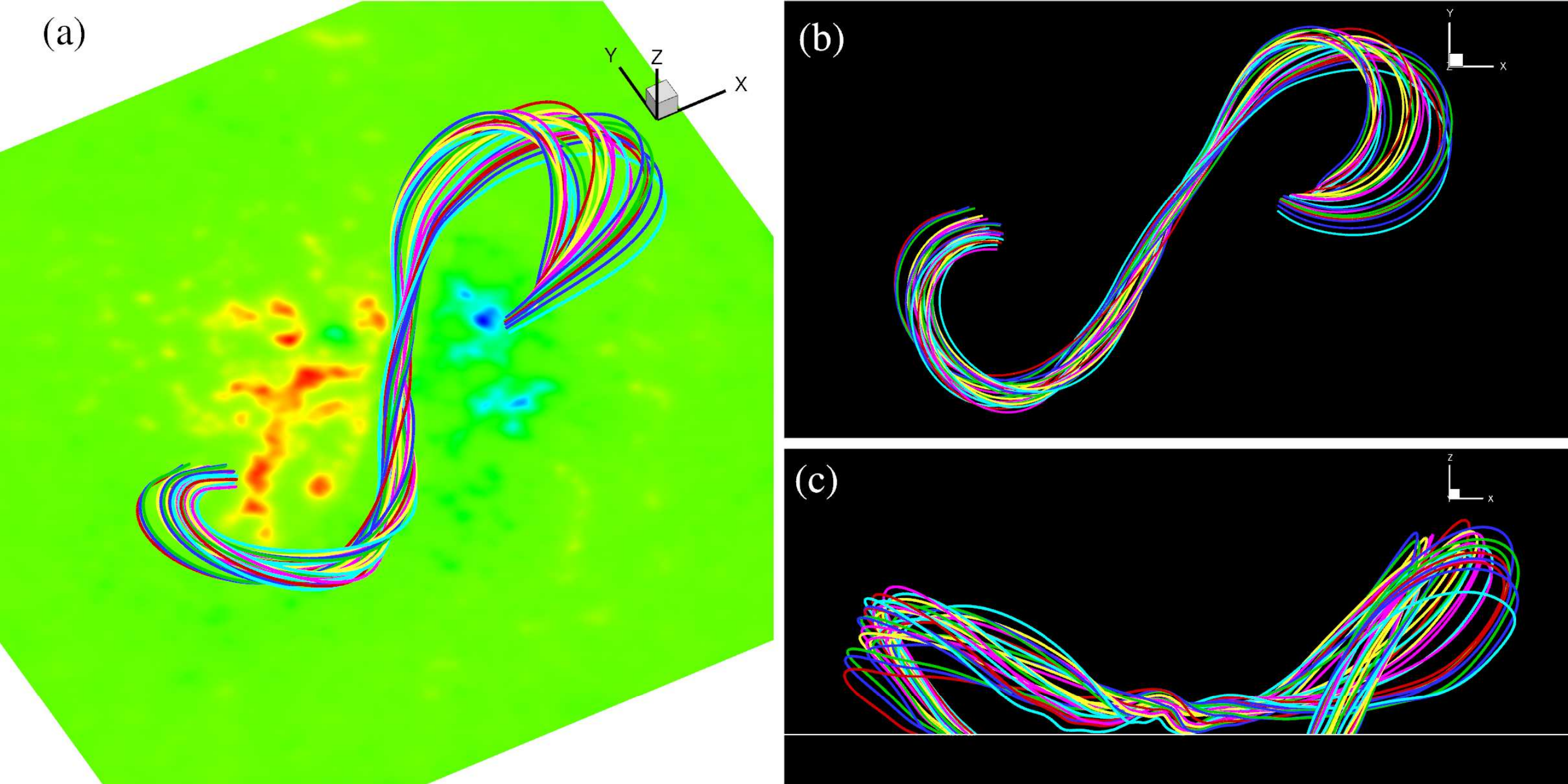}
  \caption{The van Ballegooijen reference model: different views of
    the fluxrope. (a) 3D view, (b) projection on the $x$--$y$ plane
    and (c) on the $x$--$z$ plane with the white line denoting the
    bottom. Since the middle of the fluxrope lies very close to the
    bottom, the $z$ scale in panel (c) is enlarged twice for a better
    view of the field lines.}
  \label{fig:fluxrope}
\end{figure}

We first perform the extrapolation of the chromospheric case, which
provides a largely force-free magnetogram without any
preprocessing. \Fig~\ref{fig:AA_chrom_3D} shows the 3D field lines of
extrapolation and the reference model for a side-by-side
comparison. The field lines shown are traced from foot-points equally
spaced at the bottom surface. We specially adjusted the figures to an
approximately same orientation as Paper II on purpose of visual
comparison with those results by other codes. The same field lines are
also projected on the $x$--$y$ plane in \Fig~\ref{fig:AA_chrom_xy}. As
shown from a overall view of the figures, the extrapolation reproduced
quite well the basic magnetic topology including the low-lying
S-shaped field-line bundle, and the overlying magnetic arcade that
straddles the flux rope. For a confirmation of presence of the flux
rope in the extrapolation, we select a set of field lines of the
S-shaped bundle and plot them using different colors in
\Fig~\ref{fig:fluxrope} with different perspectives. It clearly shows
that the flux rope are qualitatively recovered, encouraging us that
the code can be used to handle such relatively complex test cases. The
extrapolated flux rope is weakly twisted and its core fields basically
lies along the bottom PIL, showing a highly shear with respect to the
overlying arcades.

In Table~\ref{tab:AA_chrom} we give a quantitative evaluation and
comparison of the result with those reported in Paper II. To minimize
the (side and top) boundary effects, the metrics are applied to the
central region of $(x,y) \in [48,271]\times [48,271]$ with two
different heights: $z \in [2,61]$ for focusing on the low-lying flux
rope and $z \in [2,225]$ for including also the surrounding
potential-like field. This is done in the same way as in Paper II and
the results along with the first three preferable results in Paper II
are presented. Note that some metrics (e.g., the $C_{\rm CS}$ and
CWsin) for the reference and potential models differs slightly from
those shown in Paper II, which may be due to different precisions
(i.e., single or double precision) used in the numerical computation.

As shown by Table~\ref{tab:AA_chrom}, the extrapolation performs very
well with errors only of $\sim 2\%$ for the first two metrics. These
metrics, however, are not sensitive indicator of extrapolation
accuracy \citep{Valori2007} and thus all the methods in Paper II give
these metrics within the same level, even including the potential
field. By the more sensitive metrics $E_{\rm n}$ and $E_{\rm m}$, we
find that our result for $z\in [2,225]$ is identical with the best
result performed by Wiegelmann. For the lower height, the deviations
of these metrics with the best result are smaller than $< 2\%$ (note
that these deviations can be also partially introduced by the
different precisions in the numerical computation). This comparison
encourages us further of the success of our implementation. Our result
also gives the energy metrics very close to the reference values,
showing a good recovering of the free energy content, which is
particularly important for coronal field extrapolation. Finally the
CWsin metric behaves a bit worse than the first five metrics if
compared with the Wiegelmann's result, but it still scores better than
the following results by Wheatland and Valori.

The last two metrics $E_{\crlB}$ and $E_{\divB}$ (shown in
Table~\ref{tab:AA_divB}), both of which are non-zero but in a very low
level, mean that the reference model is very close to force-free and
divergence-free but never exact as expected (see
Section~\ref{sec:AA}). Again, our results present values very similar
to those of the reference model. Note that although both CWsin and
$E_{\crlB}$ measure the degree of force-free, the former depends more
strongly on the high-current regions \citep{Valori2007} while the
latter is not. Hence CWsin for both heights gives nearly the same
value of $0.1$, while $E_{\crlB}$ gives significant larger value
$0.06$ for the lower domain ($z\in [2,61]$) than value $0.018$ for the
full domain ($z\in [2,225]$), which shows that the residual forces are
mainly presented in the lower region. By these two metrics, we
demonstrate that our code can minimize the residual forces into a very
low values and fulfill the divergence-free condition well.

In Table~\ref{tab:AA_chrom}, the result with the multigrid-type
algorithm is also presented, which is performed on a three-grid
sequence ($80\times 80\times 64$, $160\times 160\times 128$ and the
full resolution $320\times 320\times 256$). We found that this result
(with multigrid algorithm) behaves a little bit worse compared with
the result without multigrid algorithm. This is due to the additional
numerical errors of $\crlB$ and $\divB$ introduced by interpolating
field from coarser to finer grid, which thus results in a little bit
larger values of CWsin and $\langle |f_{i}|\rangle$ than those without
multigrid algorithm. Without the multigrid algorithm, the magnetic
field splitting form can avoid such numerical errors by using a zero
$\vec B_{1}$ initially and thus performs better. However in spite of
such small disadvantage, the multigrid algorithm is still encouraged
to be adopted since its reward is a significant reduce of the
computing time (e.g., for the present test case, we find that using of
the multigrid algorithm saves approximately two-thirds of the CPU
time).

Now we present the results for the photospheric case with preprocessed
magnetogram. The preprocessing procedure can remove the net forces of
the photospheric magnetogram and hence provide the extrapolation code
a more consistent lower boundary condition than the raw
magnetogram. In Paper II, both cases with the raw and preprocessed
photospheric magnetograms are tested, and it is found that the
preprocessed case gives a significantly better result than the raw
case in which the flux rope is not reproduced at all. This
demonstrated that the preprocessing procedure is necessary for those
extrapolation codes. Besides, the smoothing involved in the
preprocessing also benefits the extrapolation code which is based on
numerical finite-difference.

\begin{figure}[\htbp]
  \centering
  \includegraphics[width=\textwidth]{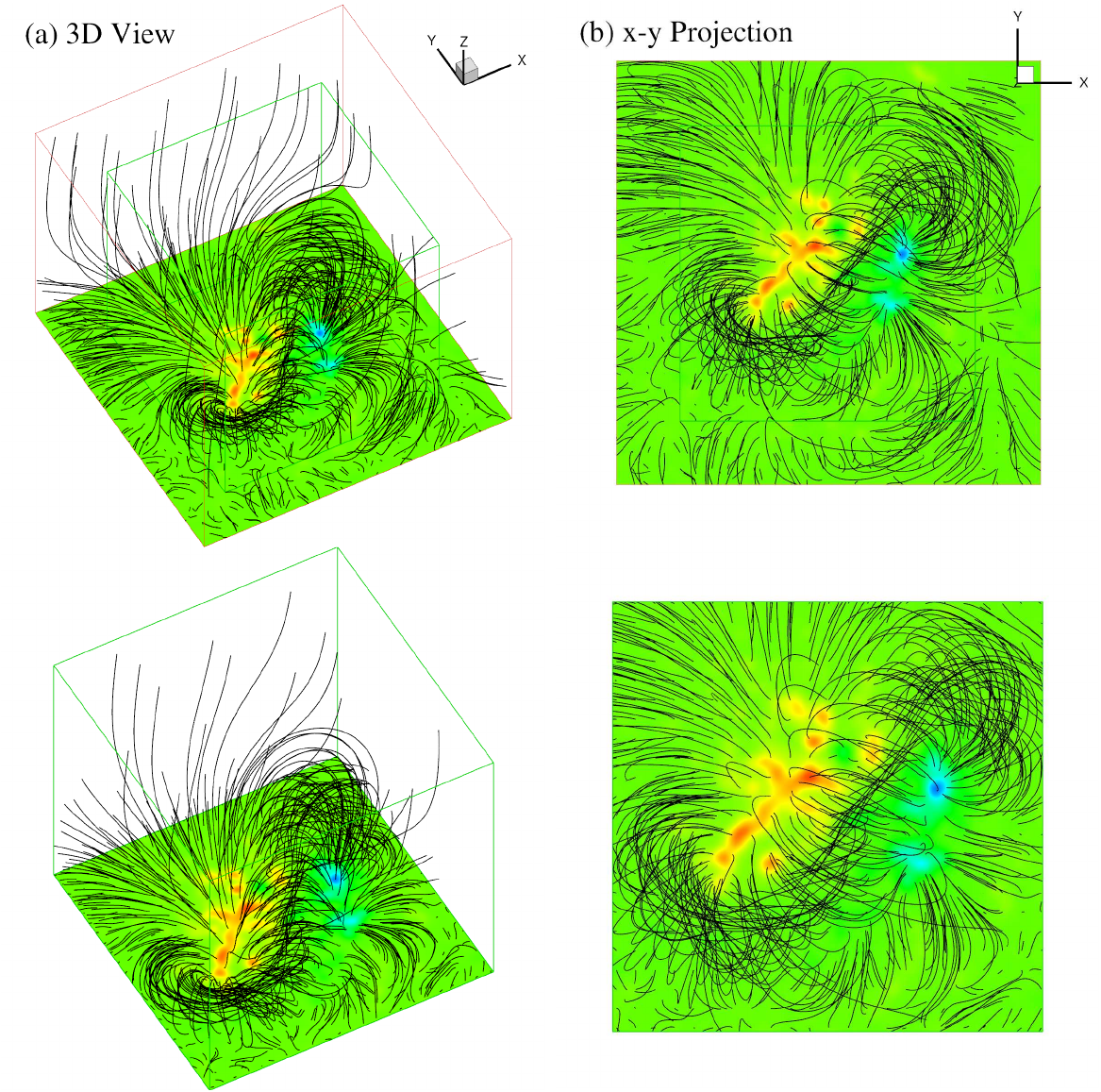}
  \caption{Photospheric test case of the van Ballegooijen reference
    model: magnetic field lines with contour of $B_{z}$ on the bottom
    surface. (a) 3D view and (b) $x$--$y$ plane projection of the
    extrapolation result. The bottom row enlarges the central region
    outlined by the small cube in the top row.}
  \label{fig:AA_photo}
\end{figure}

Our result are given in \Fig~\ref{fig:AA_photo} and
Table~\ref{tab:AA_photo}. The field lines shows that the overall
structure including the flux rope is still recovered qualitatively,
but partially. By careful comparison with the reference or the
chromospheric cases, the difference is also evident, e.g., the bundle
of S-shaped lines is much thinner in the present case, which seems
that many field lines have not reconnected fully to form a entire
S. The quantitative comparison also demonstrates that the
extrapolation quality is a bit worse than chromospheric case,
especially shown by the CWsin and $E_{\crlB}$ metrics (nearly twice of
those of the chromospheric case). It means the extrapolated field is
farther away from the exactly force-free state than the chromospheric
case. The free energy is also underestimated much, as the results in
Paper II. This is because the preprocessing does not recover a
force-free magnetogram that is entirely consistent with the reference
model. Still, it is worthwhile noting that our result again reaches
the same level of the best in Paper II, and some of the metrics,
including the most sensitive one $E_{\rm m}$, perform even better than
Wiegelmann's.

\begin{table}[htbp]
  \centering
  \begin{tabular}{llllllll}
    \hline
    \hline
    Model & $C_{\rm vec}$ & $C_{\rm CS}$ & $E_{\rm n}'$ & $E_{\rm m}'$
    & $E/E_{\rm pot}$ & CWsin & $\langle |f_{i}|\rangle$ \\ 
    \hline
    For $z \in [2,225]$ \\
    Reference          & 1.000 & 1.000 & 1.000 & 1.000 & 1.343 & 0.107 & $1.02\times 10^{-4}$\\
    {\bf Our result}   & 0.995 & 0.978 & 0.885 & 0.734 & 1.337 & 0.133 & $1.11\times 10^{-4}$ \\
    Our result$^{\rm M}$ & 0.993 & 0.975 & 0.868 & 0.712 & 1.357 & 0.142 & $1.32\times 10^{-4}$ \\
    {\bf Wiegelmann}$^{*}$ & 1.00  & 0.99  & 0.89  & 0.73  & 1.34  & 0.11  & \\
    Wheatland$^{*}$     & 0.95  & 0.98  & 0.79  & 0.70  & 1.21  & 0.15  & \\
    Valori$^{*}$        & 0.98  & 0.98  & 0.84  & 0.71  & 1.25  & 0.15  & \\ 
    Potential          & 0.852 & 0.952 & 0.687 & 0.665 & 1.000 &       & \\
    \hline
    For $z \in [2,61]$ \\
    Reference           & 1.000 & 1.000 & 1.000 & 1.000 & 1.361 & 0.103 & $1.33\times 10^{-4}$ \\
    {\bf Our result}    & 0.995 & 0.989 & 0.923 & 0.883 & 1.348 & 0.125 & $1.58\times 10^{-4}$ \\
    Our result$^{\rm M}$ & 0.993 & 0.979 & 0.902 & 0.825 & 1.366 & 0.129 &  $1.74\times 10^{-4}$\\
    {\bf Wiegelmann}$^{*}$ & 1.00  & 0.99  & 0.94  & 0.89  &   -    & 0.11  & \\
    Wheatland$^{*}$     & 0.95  & 0.95  & 0.79  & 0.76  &   -    & 0.15  & \\
    Valori$^{*}$        & 0.98  & 0.96  & 0.86  & 0.81  &   -    & 0.15  & \\
    Potential           & 0.847 & 0.901 & 0.660 & 0.678 & 1.000 &       & \\
    \hline
  \end{tabular}
  \caption{The van Ballegooijen reference model: metrics for
    extrapolation of the chromospheric case. The superscript M denotes
    the multigrid-type algorithm is used for speeding up of the
    extrapolation and the superscript $*$
    denotes the reported results in Table 3 and Table 4 of Paper II.}
  \label{tab:AA_chrom}
\end{table}

\begin{table}[htbp]
  \centering
  \begin{tabular}{lllllllll}
    \hline
    \hline
    Model & $C_{\rm vec}$ & $C_{\rm CS}$ & $E_{\rm n}'$ & $E_{\rm m}'$
    & $E/E_{\rm pot}$ & CWsin & $\langle |f_{i}|\rangle$ \\
    \hline
    For $z \in [2,225]$ \\
    Reference       & 1.000 & 1.000 & 1.000 & 1.000 & 1.531 & 0.107 & $1.02\times 10^{-4}$\\
    {\bf Our result}& 0.970 & 0.968 & 0.783 & 0.679 & 1.146 & 0.257 & $2.29\times 10^{-4}$\\
    {\bf Wiegelmann}$^{*}$ & 0.98  & 0.97  & 0.77  & 0.65  & 1.18  & 0.26  & \\
    Wheatland$^{*}$  & 0.88  & 0.96  & 0.69  & 0.65  & 1.03  & 0.11  & \\
    Potential       & 0.850 & 0.945 & 0.659 & 0.636 & 1.000 &       & \\
    \hline
    For $z \in [2,61]$ \\
    Reference       & 1.000 & 1.000 & 1.000 & 1.000 & 1.559 & 0.103 & $1.33\times 10^{-4}$\\ 
    Our result      & 0.970 & 0.980 & 0.800 & 0.791 & 1.149 & 0.230 & $3.45\times 10^{-4}$\\
    Potential       & 0.845 & 0.891 & 0.629 & 0.646 & 1.000 &       & \\
    \hline
  \end{tabular}
  \caption{The van Ballegooijen reference model: metrics for
    extrapolation of the photospheric case. The superscript $*$
    denotes the reported results in Table 3 of Paper II.}
  \label{tab:AA_photo}
\end{table}

\begin{table}[htbp]
  \centering
  \begin{tabular}{lll}
    \hline
    \hline
    Model & $E_{\crlB}$ & $E_{\divB}$ \\
    \hline
    For $z \in [2,225]$ \\
    Reference & 0.018 & 0.004\\
    Our result of the chromospheric case & 0.023 & 0.005\\
    Our result of the photospheric case  &0.048 & 0.015 \\
    \hline
    For $z \in [2,61]$ \\
    Reference & 0.060 & 0.010 \\
    Our result of the chromospheric case & 0.072 & 0.014 \\
    Our result of the photospheric case  & 0.132 & 0.040 \\
    \hline
  \end{tabular}
  \caption{The van Ballegooijen reference model: metrics of
    $E_{\crlB}$ and $E_{\divB}$.}
  \label{tab:AA_divB}
\end{table}

\section{Conclusions}
\label{sec:conclusions}

As a viable way of study magnetic field in the corona, the NLFFF
extrapolation also needs a considerable effort to be devoted for its
numerical realization. In this paper, a new numerical implementation
of NLFFF extrapolation is presented based on the MHD relaxation
method and the CESE-MHD code. Our implementation outstands for the
following aspects.

1. The magneto-frictional approach that is designed for speeding the
relaxation of the MHD system \citep{Roumeliotis1996,Valori2007} is
novelly realized by the high-performance CESE scheme on a grid without
any type of ghost zone or buffer layer.

2. The accuracy is further improved by firstly utilizing the magnetic
splitting form \citep{Tanaka1994} for NLFFF methods to totally avoid
the numerically random errors involved with the initial input.

3. Multi-method control, i.e., the diffusive and convection term, of
numerical magnetic monopoles is employed for effectively reducing the
divergence error.

4. The vector magnetogram is inputted at the bottom boundary in the
way of time-linearly modifying the potential field to matching the
magnetogram and other artificial boundaries are just fixed as the
initial potential values, which makes our implementation much easier
than other MHD relaxation methods (e.g., those by
\citet{Roumeliotis1996} and \citet{Valori2007}).

5. The code is highly parallelized with the help of PARAMESH toolkit
and performed on the share-memory parallel computer. It can be readily
realized with the AMR technique and applied to very high-resolution
magnetogram in the near future. The multigrid-type algorithm is also
incorporated into the code to speedup the computation as recommended.

We have examined the capability of the method by several reference
solutions of NLFFF that can be served as a suite of benchmark tests
for any NLFFF extrapolation code. These test cases consist of the
classic half-analytic force-free fields by Low \& Lou \citep{Low1990}
and the much more stringent and solar-like reference solution by
\citet{Ballegooijen2004}. The results show that our method are
successful and versatile for extrapolations of either the relatively
simple cases or the rather complex cases which need significant
rebuilding of the magnetic topology, e.g., the flux rope. We also
compute a suite of metrics to quantitatively analyze the results, and
shows that the solutions are extrapolated with high accuracies which
are very close to and even surpass the best results by
\citet{Wiegelmann2004} from comparison of the metrics. This
demonstrated that, at least in computation accuracy, our code performs
as good as the best one of state-of-the-art (the computing time of the
code, however, is difficult to be compared because the hardwares are
different). In addition, we introduced a pair of metrics for
assessment of the divergence-freeness and force-freeness of the
extrapolation, $E_{\divB}$ and $E_{\crlB}$, which further demonstrated
that our code can fulfill the solenoidal constraint well and minimize
the Lorentz force to the same level of the reference values.

The success of our implementation encouraged us that with a good
solver, the MHD relaxation approach can also extrapolate the NLFFF as
accurately as other good-performance algorithms, like the weighting
optimization method. This confirms again that the way of
implementation of the methods plays the same important role as their
underlying approach. It is especially worthwhile pointing out that, as
also noted by \citep{Wiegelmann2008}, the MHD relaxation approach has
a great advantage over other methods: any available time-dependent MHD
code can be adjusted for NLFFF extrapolations, which thus saves the
major effort that should be made to develop a new code from scratch
for a special method. We are also inspired by \citet{Valori2007}, who
show that a higher order scheme can significantly advance the
extrapolation. In our project, an arbitrary high order CESE scheme is
under development and is expected to be used for future improvement of
our implementation.

Recently, more critical tests of extrapolation codes have been
performed by \citet{Schrijver2008} and \citet{Derosa2009} based on
vector magnetograms of AR 10930 and 10953 from Hinode/SOT and observed
coronal loops. It is found that the Grad-Rubin-style current-field
iteration implemented by \citet{Wheatland2006} surpassed the
\citet{Wiegelmann2004}'s code that performs best in the benchmark
tests, and basically the results by different methods are very
inconsistent with each other. This shows that the idealized tests are
unable to assess the code's ability to deal with various uncertainties
or errors in the real magnetograms and more critical assessment of the
code using realistic vector magnetograms is also planned in our future
work.

The present extrapolation in Cartesian geometry is often limited to
relatively local areas, e.g., a single active region without any
relationship with others. However, the active regions cannot be
isolated since they generally interact with neighboring ARs or
overlaying large scale fields. It is also pointed out that the fields
of view in Cartesian box are often too small to properly characterize
the entire relevant current system \citep{Derosa2009}. To study the
connectivity between multi-active regions and extrapolate in a larger
field of view, it is necessary to take into account the curvature of
the Sun's surface by extrapolation in spherical geometry partly or
even entirely, i.e., including the global corona
\citep{Wiegelmann2007,Tadesse2011,Tadesse2011A}. Moreover, a global
NLFFF extrapolation can also avoid any lateral artificial boundaries
which are inescapable and cause issues in Cartesian codes. We are now
on the way of developing a global NLFFF code for the new era of
routinely observation of the global vector magnetogram (which will be
opened by SDO/HMI). Recently in a project of constructing a
data-driven MHD model for the global coronal evolution, we have
established the CESE method on a so-called Yin-Yang overlapping grid
in spherical geometry \citep{Kageyama2004}. This implementation,
combined with our present NLFFF code, will make the realization of a
global NLFFF extrapolation very viable if provided with the global
vector magnetogram. The Yin-Yang grid is composed of two identical
component grids that are combined in a complemental way to cover an
entire spherical surface with partial overlap on their
boundaries. Each component grid is a low latitude part of the
latitude-longitude grid without the pole and hence the grid spacing on
the sphere surface is quasi-uniform. In this way, we can avoid the
problem of grid convergence or grid singularity at both poles which
will otherwise arise if an entire spherical-coordinate grid is used,
as \citet{Wiegelmann2007} has pointed out. However, up to now, there
is no suitable test case used for the global NLFFF extrapolation other
than the simple axially-symmetric Low \& Lou cases. Most recently,
\citet{Contopoulos2011} give a variety of global near force-free
solutions by a force-free electrodynamics code, using solely the
radial magnetogram. Their solutions, however, are not unique to the
same radial magnetogram, but depends on the initial conditions and on
the particular approach to steady-state \footnote{They are improving
  their code to also make use of the vector magnetogram to define the
  solutions uniquely (by private communications)}. Anyway, we believe
their solutions can be used as much more realistic and solar-like
tests for global NLFFF codes than the semi-analytic solutions. In our
future work, we will develop and test a new global NLFFF code by the
global force-free solutions from \citep{Contopoulos2011}.

\acknowledgments 

The work is jointly supported by the National Natural Science
Foundation of China (41031066, 40921063, 40890162, and 41074122), the
973 project under grant 2012CB825601, and the Specialized Research
Fund for State Key Laboratories. We are grateful to Dr. Marc Derosa
for providing the data of the van Ballegooijen reference model. The
PARAMESH software used in this work was developed at the NASA Goddard
Space Flight Center and Drexel University under NASA's HPCC and
ESTO/CT Projects and under Grant NNG04GP79G from the NASA/AISR
project.

\appendix

\section{The specific form of \Eq~(\ref{eq:main})}

\begin{eqnarray}
  \mathbf{F} =
  \left(\begin{array}{c}
      \noalign{\medskip}\vec B_{1}\cdot\vec B_{1}/2-B_{1x}^{2}-(B_{0x}B_{1x}+B_{1x}B_{0x})+\vec B_{0}\cdot\vec B_{1}\\
      \noalign{\medskip}-B_{1x}B_{1y}-(B_{0x}B_{1y}+B_{1x}B_{0y})\\
      \noalign{\medskip}-B_{1x}B_{1z}-(B_{0x}B_{1z}+B_{1x}B_{0z})\\
      \noalign{\medskip}0\\
      \noalign{\medskip}v_{x}B_{y}-v_{y}B_{x}\\
      \noalign{\medskip}v_{x}B_{z}-v_{z}B_{x}\\
      \noalign{\medskip}0\\
      \noalign{\medskip}0\\
      \noalign{\medskip}0
    \end {array} \right);
  \nonumber\\
  \mathbf{G} =
   \left(\begin{array}{c}
       \noalign{\medskip}-B_{1y}B_{1x}-(B_{0y}B_{1x}+B_{1y}B_{0x})\\
      \noalign{\medskip}\vec B_{1}\cdot\vec B_{1}/2-B_{1y}^{2}-(B_{0y}B_{1y}+B_{1y}B_{0y})+\vec B_{0}\cdot\vec B_{1}\\
      \noalign{\medskip}-B_{1y}B_{1z}-(B_{0y}B_{1z}+B_{1y}B_{0z})\\
      \noalign{\medskip}v_{y}B_{x}-v_{x}B_{y}\\
      \noalign{\medskip}0\\
      \noalign{\medskip}v_{y}B_{z}-v_{z}B_{y}\\
      \noalign{\medskip}0\\
      \noalign{\medskip}0\\
      \noalign{\medskip}0
    \end {array} \right);  
  \nonumber\\
   \mathbf{H} =
  \left(\begin{array}{c}
      \noalign{\medskip}-B_{1z}B_{1x}-(B_{0z}B_{1x}+B_{1z}B_{0x})\\
      \noalign{\medskip}-B_{1z}B_{1y}-(B_{0z}B_{1y}+B_{1z}B_{0y})\\
      \noalign{\medskip}\vec B_{1}\cdot\vec B_{1}/2-B_{1z}^{2}-(B_{0z}B_{1z}+B_{1z}B_{0z})+\vec B_{0}\cdot\vec B_{1}\\
      \noalign{\medskip}v_{z}B_{x}-v_{x}B_{z}\\
      \noalign{\medskip}v_{z}B_{y}-v_{y}B_{z}\\
      \noalign{\medskip}0\\
      \noalign{\medskip}0\\
      \noalign{\medskip}0\\
      \noalign{\medskip}0
    \end {array} \right);
\end{eqnarray}

\begin{eqnarray}
  \vec F_{\nu} = (0,0,0,\mu\divB_{1},0,0,0,0,0)^{T};\nonumber\\
  \vec G_{\nu} = (0,0,0,0,\mu\divB_{1},0,0,0,0)^{T};\nonumber\\
  \vec H_{\nu} = (0,0,0,0,0,\mu\divB_{1},0,0,0)^{T};
\end{eqnarray}

\begin{equation}
  \vec S = (-\nu\rho v_{x},-\nu\rho v_{y},-\nu\rho
  v_{z},v_{x}\divB_{1},v_{y}\divB_{1},v_{z}\divB_{1},0,0,0)
\end{equation}


\newpage

\newpage

\begin{thebibliography}{72}
\expandafter\ifx\csname natexlab\endcsname\relax\def\natexlab#1{#1}\fi

\bibitem[{{Abbett} \& {Fisher}(2003)}]{Abbett2003}
{Abbett}, W.~P. \& {Fisher}, G.~H. 2003, \apj, 582, 475

\bibitem[{{Abbett} {et~al.}(2004){Abbett}, {Miki{\'c}}, {Linker}, {McTiernan},
  {Magara}, \& {Fisher}}]{Abbett2004}
{Abbett}, W.~P., {Miki{\'c}}, Z., {Linker}, J.~A., {McTiernan}, J.~M.,
  {Magara}, T., \& {Fisher}, G.~H. 2004, J. Atmos. Sol.-Terr. Phys., 66, 1257

\bibitem[{{Altschuler} \& {Newkirk}(1969)}]{Altschuler1969}
{Altschuler}, M.~D. \& {Newkirk}, G. 1969, \solphys, 9, 131

\bibitem[{{Aly}(1989)}]{Aly1989}
{Aly}, J.~J. 1989, \solphys, 120, 19

\bibitem[{{Amari} {et~al.}(1997){Amari}, {Aly}, {Luciani}, {Boulmezaoud}, \&
  {Mikic}}]{Amari1997}
{Amari}, T., {Aly}, J.~J., {Luciani}, J.~F., {Boulmezaoud}, T.~Z., \& {Mikic},
  Z. 1997, \solphys, 174, 129

\bibitem[{{Amari} {et~al.}(2006){Amari}, {Boulmezaoud}, \& {Aly}}]{Amari2006}
{Amari}, T., {Boulmezaoud}, T.~Z., \& {Aly}, J.~J. 2006, \aap, 446, 691

\bibitem[{{Amari} {et~al.}(1999){Amari}, {Boulmezaoud}, \& {Mikic}}]{Amari1999}
{Amari}, T., {Boulmezaoud}, T.~Z., \& {Mikic}, Z. 1999, \aap, 350, 1051

\bibitem[{{Aschwanden}(2005)}]{Aschwanden2005}
{Aschwanden}, M.~J. 2005, {Physics of the Solar Corona. An Introduction with
  Problems and Solutions (2nd edition)}, ed. {Aschwanden, M.~J.}

\bibitem[{{Canou} \& {Amari}(2010)}]{Canou2010}
{Canou}, A. \& {Amari}, T. 2010, \apj, 715, 1566

\bibitem[{Chang(2002)}]{Chang2002}
Chang, S.~C. 2002, AIAA Paper, 2002-3890

\bibitem[{{Chodura} \& {Schlueter}(1981)}]{Chodura1981}
{Chodura}, R. \& {Schlueter}, A. 1981, \jcphy, 41, 68

\bibitem[{{Contopoulos} {et~al.}(2011){Contopoulos}, {Kalapotharakos}, \&
  {Georgoulis}}]{Contopoulos2011}
{Contopoulos}, I., {Kalapotharakos}, C., \& {Georgoulis}, M.~K. 2011, \solphys,
  269, 351

\bibitem[{{Dedner} {et~al.}(2002){Dedner}, {Kemm}, {Kr{\"o}ner}, {Munz},
  {Schnitzer}, \& {Wesenberg}}]{Dedner2002}
{Dedner}, A., {Kemm}, F., {Kr{\"o}ner}, D., {Munz}, C., {Schnitzer}, T., \&
  {Wesenberg}, M. 2002, \jcphy, 175, 645

\bibitem[{{Dellar}(2001)}]{Dellar2001}
{Dellar}, P.~J. 2001, \jcphy, 172, 392

\bibitem[{{DeRosa} {et~al.}(2009){DeRosa}, {Schrijver}, {Barnes}, {Leka},
  {Lites}, {Aschwanden}, {Amari}, {Canou}, {McTiernan}, {R{\'e}gnier},
  {Thalmann}, {Valori}, {Wheatland}, {Wiegelmann}, {Cheung}, {Conlon},
  {Fuhrmann}, {Inhester}, \& {Tadesse}}]{Derosa2009}
{DeRosa}, M.~L., {Schrijver}, C.~J., {Barnes}, G., {Leka}, K.~D., {Lites},
  B.~W., {Aschwanden}, M.~J., {Amari}, T., {Canou}, A., {McTiernan}, J.~M.,
  {R{\'e}gnier}, S., {Thalmann}, J.~K., {Valori}, G., {Wheatland}, M.~S.,
  {Wiegelmann}, T., {Cheung}, M.~C.~M., {Conlon}, P.~A., {Fuhrmann}, M.,
  {Inhester}, B., \& {Tadesse}, T. 2009, \apj, 696, 1780

\bibitem[{{Feng} {et~al.}(2006){Feng}, {Hu}, \& {Wei}}]{Feng2006}
{Feng}, X., {Hu}, Y., \& {Wei}, F. 2006, \solphys, 235, 235

\bibitem[{{Feng} {et~al.}(2007){Feng}, {Zhou}, \& {Wu}}]{Feng2007}
{Feng}, X., {Zhou}, Y., \& {Wu}, S.~T. 2007, \apj, 655, 1110

\bibitem[{{Feng} {et~al.}(2010){Feng}, {Yang}, {Xiang}, {Wu}, {Zhou}, \&
  {Zhong}}]{Feng2010}
{Feng}, X.~S., {Yang}, L.~P., {Xiang}, C.~Q., {Wu}, S.~T., {Zhou}, Y.~F., \&
  {Zhong}, D.~K. 2010, \apj, 723, 300

\bibitem[{Grad \& Rubin(1958)}]{Grad1958}
Grad, H. \& Rubin, H. 1958, in 2nd Int. Conf. Peac. Uses of Atom. Energy,
  Vol.~31, 386

\bibitem[{{Hayashi}(2005)}]{Hayashi2005}
{Hayashi}, K. 2005, \apjs, 161, 480

\bibitem[{{He} \& {Wang}(2008)}]{He2008}
{He}, H. \& {Wang}, H. 2008, \jgr, 113, 5

\bibitem[{{He} {et~al.}(2011){He}, {Wang}, \& {Yan}}]{He2011}
{He}, H., {Wang}, H., \& {Yan}, Y. 2011, \jgr, 116, 1101

\bibitem[{{Hoeksema}(1984)}]{Hoeksema1984}
{Hoeksema}, J.~T. 1984, PhD thesis, Stanford Univ., CA.

\bibitem[{{Inhester} \& {Wiegelmann}(2006)}]{Inhester2006}
{Inhester}, B. \& {Wiegelmann}, T. 2006, \solphys, 235, 201

\bibitem[{{Inoue} {et~al.}(2011){Inoue}, {Kusano}, {Magara}, {Shiota}, \&
  {Yamamoto}}]{Inoue2011}
{Inoue}, S., {Kusano}, K., {Magara}, T., {Shiota}, D., \& {Yamamoto}, T.~T.
  2011, \apj, 738, 161

\bibitem[{{Jiang} {et~al.}(2011){Jiang}, {Feng}, {Fan}, \& {Xiang}}]{Jiang2011}
{Jiang}, C.~W., {Feng}, X.~S., {Fan}, Y.~L., \& {Xiang}, C.~Q. 2011, \apj, 727, 101

\bibitem[{Jiang {et~al.}(2012)Jiang, Cui, \& Feng}]{Jiang2011s}
Jiang, C.~W., Cui, S.~X., \& Feng, X.~S. 2012, Computers \& Fluids, 54, 105

\bibitem[{Jiang {et~al.}(2010)Jiang, Feng, Zhang, \& Zhong}]{Jiang2010}
Jiang, C.~W., Feng, X.~S., Zhang, J., \& Zhong, D.~K. 2010, \solphys, 267,
  463

\bibitem[{{Kageyama} \& {Sato}(2004)}]{Kageyama2004}
{Kageyama}, A. \& {Sato}, T. 2004, Geochemistry, Geophysics, Geosystems, 5,
  9005

\bibitem[{{Linker} {et~al.}(1999){Linker}, {Miki{\'c}}, {Biesecker}, {Forsyth},
  {Gibson}, {Lazarus}, {Lecinski}, {Riley}, {Szabo}, \&
  {Thompson}}]{Linker1999}
{Linker}, J.~A., {Miki{\'c}}, Z., {Biesecker}, D.~A., {Forsyth}, R.~J.,
  {Gibson}, S.~E., {Lazarus}, A.~J., {Lecinski}, A., {Riley}, P., {Szabo}, A.,
  \& {Thompson}, B.~J. 1999, \jgr, 104, 9809

\bibitem[{{Low} \& {Lou}(1990)}]{Low1990}
{Low}, B.~C. \& {Lou}, Y.~Q. 1990, \apj, 352, 343

\bibitem[{{MacNeice} {et~al.}(2000){MacNeice}, {Olson}, {Mobarry}, {de
  Fainchtein}, \& {Packer}}]{MacNeice2000}
{MacNeice}, P., {Olson}, K.~M., {Mobarry}, C., {de Fainchtein}, R., \&
  {Packer}, C. 2000, Comput. Phys. Commun., 126, 330

\bibitem[{{Marder}(1987)}]{Marder1987}
{Marder}, B. 1987, \jcphy, 68, 48

\bibitem[{{McClymont} {et~al.}(1997){McClymont}, {Jiao}, \&
  {Mikic}}]{McClymont1997}
{McClymont}, A.~N., {Jiao}, L., \& {Mikic}, Z. 1997, \solphys, 174, 191

\bibitem[{Metcalf {et~al.}(2008)Metcalf, DeRosa, Schrijver, Barnes, van
  Ballegooijen, Wiegelmann, Wheatland, Valori, \& McTtiernan}]{Metcalf2008}
Metcalf, T.~R., DeRosa, M.~L., Schrijver, C.~J., Barnes, G., van Ballegooijen,
  A.~A., Wiegelmann, T., Wheatland, M.~S., Valori, G., \& McTtiernan, J.~M.
  2008, \solphys, 247, 269

\bibitem[{{Miki{\'c}} {et~al.}(1999){Miki{\'c}}, {Linker}, {Schnack},
  {Lionello}, \& {Tarditi}}]{Mikic1999}
{Miki{\'c}}, Z., {Linker}, J.~A., {Schnack}, D.~D., {Lionello}, R., \&
  {Tarditi}, A. 1999, Phys. Plasmas, 6, 2217

\bibitem[{{Mikic} \& {McClymont}(1994)}]{Mikic1994a}
{Mikic}, Z. \& {McClymont}, A.~N. 1994, in Astronomical Society of the Pacific
  Conference Series, Vol.~68, Solar Active Region Evolution: Comparing Models
  with Observations, ed. {K.~S.~Balasubramaniam \& G.~W.~Simon}, 225--+

\bibitem[{{Nakagawa}(1974)}]{Nakagawa1974}
{Nakagawa}, Y. 1974, \apj, 190, 437

\bibitem[{{Nakamizo} {et~al.}(2009){Nakamizo}, {Tanaka}, {Kubo}, {Kamei},
  {Shimazu}, \& {Shinagawa}}]{Nakamizo2009}
{Nakamizo}, A., {Tanaka}, T., {Kubo}, Y., {Kamei}, S., {Shimazu}, H., \&
  {Shinagawa}, H. 2009, \jgr (Space Phys.), 114,
  A07109

\bibitem[{{Powell} {et~al.}(1999){Powell}, {Roe}, {Linde}, {Gombosi}, \& {de
  Zeeuw}}]{Powell1999}
{Powell}, K.~G., {Roe}, P.~L., {Linde}, T.~J., {Gombosi}, T.~I., \& {de Zeeuw},
  D.~L. 1999, \jcphy, 154, 284

\bibitem[{{Roumeliotis}(1996)}]{Roumeliotis1996}
{Roumeliotis}, G. 1996, \apj, 473, 1095

\bibitem[{Sakurai(1981)}]{Sakurai1981}
Sakurai, T. 1981, \solphys, 69, 343

\bibitem[{{Sakurai}(1989)}]{Sakurai1989}
{Sakurai}, T. 1989, \ssr, 51, 11

\bibitem[{{Schrijver} {et~al.}(2006){Schrijver}, {De Rosa}, {Metcalf}, {Liu},
  {McTiernan}, {R{\'e}gnier}, {Valori}, {Wheatland}, \&
  {Wiegelmann}}]{Schrijver2006}
{Schrijver}, C.~J., {De Rosa}, M.~L., {Metcalf}, T.~R., {Liu}, Y., {McTiernan},
  J., {R{\'e}gnier}, S., {Valori}, G., {Wheatland}, M.~S., \& {Wiegelmann}, T.
  2006, \solphys, 235, 161

\bibitem[{{Schrijver} {et~al.}(2008){Schrijver}, {DeRosa}, {Metcalf}, {Barnes},
  {Lites}, {Tarbell}, {McTiernan}, {Valori}, {Wiegelmann}, {Wheatland},
  {Amari}, {Aulanier}, {D{\'e}moulin}, {Fuhrmann}, {Kusano}, {R{\'e}gnier}, \&
  {Thalmann}}]{Schrijver2008}
{Schrijver}, C.~J., {DeRosa}, M.~L., {Metcalf}, T., {Barnes}, G., {Lites}, B.,
  {Tarbell}, T., {McTiernan}, J., {Valori}, G., {Wiegelmann}, T., {Wheatland},
  M.~S., {Amari}, T., {Aulanier}, G., {D{\'e}moulin}, P., {Fuhrmann}, M.,
  {Kusano}, K., {R{\'e}gnier}, S., \& {Thalmann}, J.~K. 2008, \apj, 675, 1637

\bibitem[{{Song} {et~al.}(2006){Song}, {Fang}, {Tang}, {Wu}, \&
  {Zhang}}]{Song2006}
{Song}, M.~T., {Fang}, C., {Tang}, Y.~H., {Wu}, S.~T., \& {Zhang}, Y.~A. 2006,
  \apj, 649, 1084

\bibitem[{{Tadesse} {et~al.}(2011{\natexlab{a}}){Tadesse}, {Wiegelmann},
  {Inhester}, \& {Pevtsov}}]{Tadesse2011}
{Tadesse}, T., {Wiegelmann}, T., {Inhester}, B., \& {Pevtsov}, A.
  2011{\natexlab{a}}, \solphys, 167

\bibitem[{{Tadesse} {et~al.}(2011{\natexlab{b}}){Tadesse}, {Wiegelmann},
  {Inhester}, \& {Pevtsov}}]{Tadesse2011A}
---. 2011{\natexlab{b}}, \aap, 527, A30

\bibitem[{{Tanaka}(1994)}]{Tanaka1994}
{Tanaka}, T. 1994, \jcphy, 111, 381

\bibitem[{{T{\'o}th} \& {Roe}(2002)}]{Toth2002}
{T{\'o}th}, G. \& {Roe}, P.~L. 2002, \jcphy, 180, 736

\bibitem[{{Valori} {et~al.}(2007){Valori}, {Kliem}, \& {Fuhrmann}}]{Valori2007}
{Valori}, G., {Kliem}, B., \& {Fuhrmann}, M. 2007, \solphys, 245, 263

\bibitem[{{Valori} {et~al.}(2005){Valori}, {Kliem}, \& {Keppens}}]{Valori2005}
{Valori}, G., {Kliem}, B., \& {Keppens}, R. 2005, \aap, 433, 335

\bibitem[{{van Ballegooijen}(2004)}]{Ballegooijen2004}
{van Ballegooijen}, A.~A. 2004, \apj, 612, 519

\bibitem[{{van Ballegooijen} {et~al.}(2007){van Ballegooijen}, {Deluca},
  {Squires}, \& {Mackay}}]{Ballegooijen2007}
{van Ballegooijen}, A.~A., {Deluca}, E.~E., {Squires}, K., \& {Mackay}, D.~H.
  2007, J. Atmos. Sol.-Terr. Phys., 69, 24

\bibitem[{{van Ballegooijen} {et~al.}(2000){van Ballegooijen}, {Priest}, \&
  {Mackay}}]{Ballegooijen2000}
{van Ballegooijen}, A.~A., {Priest}, E.~R., \& {Mackay}, D.~H. 2000, \apj, 539,
  983

\bibitem[{{Wang} {et~al.}(2011){Wang}, {Wu}, {Tandberg-Hanssen}, \&
  {Hill}}]{Wang2011}
{Wang}, A.~H., {Wu}, S.~T., {Tandberg-Hanssen}, E., \& {Hill}, F. 2011, \apj,
  732, 19

\bibitem[{{Welsch} {et~al.}(2004){Welsch}, {Fisher}, {Abbett}, \&
  {Regnier}}]{Welsch2004}
{Welsch}, B.~T., {Fisher}, G.~H., {Abbett}, W.~P., \& {Regnier}, S. 2004, \apj,
  610, 1148

\bibitem[{{Wheatland}(2004)}]{Wheatland2004}
{Wheatland}, M.~S. 2004, \solphys, 222, 247

\bibitem[{{Wheatland}(2006)}]{Wheatland2006}
---. 2006, \solphys, 238, 29

\bibitem[{{Wheatland} {et~al.}(2000){Wheatland}, {Sturrock}, \&
  {Roumeliotis}}]{Wheatland2000}
{Wheatland}, M.~S., {Sturrock}, P.~A., \& {Roumeliotis}, G. 2000, \apj, 540,
  1150

\bibitem[{{Wiegelmann}(2004)}]{Wiegelmann2004}
{Wiegelmann}, T. 2004, \solphys, 219, 87

\bibitem[{{Wiegelmann}(2007)}]{Wiegelmann2007}
---. 2007, \solphys, 240, 227

\bibitem[{{Wiegelmann}(2008)}]{Wiegelmann2008}
---. 2008, \jgr, 113, 3

\bibitem[{{Wiegelmann} \& {Neukirch}(2006)}]{Wiegelmann2006}
{Wiegelmann}, T. \& {Neukirch}, T. 2006, \aap, 457, 1053

\bibitem[{{Wu} {et~al.}(1990){Wu}, {Sun}, {Chang}, {Hagyard}, \&
  {Gary}}]{Wu1990}
{Wu}, S.~T., {Sun}, M.~T., {Chang}, H.~M., {Hagyard}, M.~J., \& {Gary}, G.~A.
  1990, \apj, 362, 698

\bibitem[{{Wu} {et~al.}(2009){Wu}, {Wang}, {Gary}, {Kucera}, {Rybak}, {Liu},
  {Vr{\'s}nak}, \& {Yurchyshyn}}]{Wu2009}
{Wu}, S.~T., {Wang}, A.~H., {Gary}, G.~A., {Kucera}, A., {Rybak}, J., {Liu},
  Y., {Vr{\'s}nak}, B., \& {Yurchyshyn}, V. 2009, Adv. Space Res.,
  44, 46

\bibitem[{{Wu} {et~al.}(2006){Wu}, {Wang}, {Liu}, \& {Hoeksema}}]{Wu2006}
{Wu}, S.~T., {Wang}, A.~H., {Liu}, Y., \& {Hoeksema}, J.~T. 2006, \apj, 652,
  800

\bibitem[{{Wu} {et~al.}(2001){Wu}, {Zheng}, {Wang}, {Thompson}, {Plunkett},
  {Zhao}, \& {Dryer}}]{Wu2001}
{Wu}, S.~T., {Zheng}, H., {Wang}, S., {Thompson}, B.~J., {Plunkett}, S.~P.,
  {Zhao}, X.~P., \& {Dryer}, M. 2001, \jgr, 106, 25089

\bibitem[{{Yan} \& {Li}(2006)}]{Yan2006}
{Yan}, Y. \& {Li}, Z. 2006, \apj, 638, 1162

\bibitem[{{Yan} \& {Sakurai}(2000)}]{Yan2000}
{Yan}, Y. \& {Sakurai}, T. 2000, \solphys, 195, 89

\bibitem[{{Yang} {et~al.}(1986){Yang}, {Sturrock}, \& {Antiochos}}]{Yang1986}
{Yang}, W.~H., {Sturrock}, P.~A., \& {Antiochos}, S.~K. 1986, \apj, 309, 383

\bibitem[{{Zhang} {et~al.}(2006){Zhang}, {John Yu}, {Henry Lin}, {Chang}, \&
  {Blankson}}]{Zhang2006}
{Zhang}, M., {John Yu}, S., {Henry Lin}, S., {Chang}, S., \& {Blankson}, I.
  2006, \jcphy, 214, 599

\end{thebibliography}

\end{document}